\documentclass[twocolumn,floatfix]{aastex631}
\pdfoutput=1
\usepackage{amsmath,amstext,multirow,txfonts}
\usepackage[T1]{fontenc}
\usepackage{xcolor}
\usepackage{hyperref}
\usepackage[figure,figure*]{hypcap}

\def\be{\begin{eqnarray}}   \def\ee{\end{eqnarray}}
\def\ben{\begin{equation}\begin{aligned}} \def\een{\end{aligned}\end{equation}}
\shorttitle{Motion in stellar halo}
\shortauthors{Sharma et al.}
\newcommand{\kpc}{\>{\rm kpc}}

\newcommand{\kms}{\>{\rm km}\,{\rm s}^{-1}}

\usepackage{ulem}

\begin{document}

\title{Can radial motions in the stellar halo constrain the rate of change of mass in the Galaxy?}
\shorttitle{Radial motions in the stellar halo}
\author[0000-0002-0920-809X]{Sanjib Sharma}
\affiliation{Sydney Institute for Astronomy, School of Physics, The University of Sydney, NSW 2006, Australia}
\affiliation{ARC Centre of Excellence for All Sky Astrophysics in Three Dimensions (ASTRO-3D), Australia}

\author[0000-0001-7516-4016]{Joss Bland-Hawthorn}
\affiliation{Sydney Institute for Astronomy, School of Physics, The University of Sydney, NSW 2006, Australia}
\affiliation{ARC Centre of Excellence for All Sky Astrophysics in Three Dimensions (ASTRO-3D), Australia}

\author[0000-0002-1566-8148]{Joseph Silk}
\affiliation{Institut d'Astrophysique de Paris, 98 bis Bd Arago, Paris, France}

\author[0000-0002-5074-9998]{Celine Boehm}
\affiliation{School of Physics, The University of Sydney, NSW 2006, Australia}
\affiliation{ARC Centre of Excellence for Dark Matter Particle Physics, Australia}

\begin{abstract}
A change in the mass of the Galaxy with time will leave its imprint on the motions of the stars, with stars having radially outward (mass loss) or inward (mass accretion) bulk motions.
Here we test the feasibility of using the mean radial motion of stars in the stellar halo to
constrain the rate of change of mass in the Galaxy, for example, due to decay of dark matter into  invisible dark sector particles
or more conservatively from the settling of baryons.
In the current $\Lambda$CDM paradigm of structure formation, the stellar halo is formed by accretion of
satellites onto the host galaxy. Over time, as the satellites
disrupt and phase mix, the mean radial motion $\langle V_{R}\rangle$ of the stellar halo is eventually expected to be close to zero.
But most halos have substructures due to
incomplete mixing of specific accretion events and this can
lead to nonzero $\langle V_{R}\rangle$ in them.
Using simulations,
we measure the mean radial motion,  $\langle V_{R}\rangle$, of stars in 13 $\Lambda$CDM stellar halos lying in a spherical shell of radius 30 kpc. We find that for most halos, the shell motion is quite
small, with 75\% of halos having $\langle V_{R}\rangle \lesssim 1.2 \kms$.
When substructures are removed by using a clustering algorithm, $\langle V_{R}\rangle$ is reduced even further, with 75\% of halos having $\langle V_{R}\rangle \lesssim 0.6$ km s$^{-1}$.
A value of $\langle V_{R}\rangle \approx 0.6 \kms$ can be attained corresponding to a  galactic mass loss rate of 2\% per Gyr.
We show that this can place constraints on
dark matter decay parameters such as the decay lifetime
and the kick velocity that is imparted to the daughter particle. The advent of all-sky stellar surveys involving millions to billions of stars is encouraging for detecting signatures of dark matter decay.
\end{abstract}

\section{Introduction}
Dark matter dominates the outer galaxy but may not be absolutely stable. At the very least, this has to be demonstrated to within observational limits. Dark matter is generally considered to consist of weakly interacting particles that are hitherto undetected. Search for observational signatures is a major industry \citep{2010ARA&A..48..495F}, including  deep underground direct searches \citep{2017NatPh..13..212L}, indirect searches in astronomical systems, including the Universe itself, and high energy particle accelerator searches at the LHC \citep{2017IJMPA..3230006K} and elsewhere. 

The most popular sought-after signals  typically involved  self-annihilation of  heavy neutral  particles into charged Standard Model particles
\citep[see e.g.,][for the original idea]{Silk:1984zy}.
However many other avenues have been considered as early as in the 1980s, including the decay of dark matter particles, e.g.,  \citet{Dicus:1977qy,Cabibbo:1981er} and \citet{Ellis:1984eq}. Decaying dark matter scenarios gained further traction in the past two decades after puzzling excesses in cosmic ray and X-ray observations emerged, see e.g.,  \citet{Chen:2008qs,Ibarra:2008jk,Yin:2008bs}, or for more modern references \citet{ 2020JCAP...08..035V,2022arXiv220306508C}, as well as   \citet{Boyarsky:2014ska,Jeltema:2014qfa,Riemer-Sorensen:2014yda} and references therein. As the injection of charged particles in dark matter halos and our cosmic neighbourhood could lead to excess in cosmic ray, neutrinos, X-ray, gamma-ray and radio spectra, these could be used to set strong limits on the dark matter characteristics and, in particular, constrain its mass vs interaction strength and therefore lifetime.

More recently however it was suggested that dark matter could decay or annihilate into a dark (possibly secluded) sector. While such scenarios would be impossible to detect by traditional means, \cite{2008MNRAS.388.1869A,2010PhRvD..82l3521P,2014MNRAS.445..614W} showed that their impact on the number of satellite companions of the Milky Way would provide nonetheless a way to test their validity. More recently,
by comparing cosmological simulations of decaying dark matter with the observed Milky Way satellite population \citet{2022arXiv220111740M} were  able to place constraints on the decay lifetime and the associated kick velocity  
Here we go a step further and examine whether the invisible dark matter decay would also affect galaxy dynamics and provide  complementary limits to previous works, including \citet{2022arXiv220505636A}.

These questions are  more than academic. The Hubble tension (discrepancy between the local measurement of Hubble constant with that from the cosmic microwave background) has reinvigorated discussions about the possible instability of   dark energy \citep{2019PhRvL.122v1301P,2021PhRvD.104l3550P} and dark matter \citep{2020JCAP...07..026P, 2021EPJC...81..954F}, but see also \citet{2022PhRvD.105j3512A}. Models based on either of these hypotheses have the potential of  reducing the Hubble tension by modifying  the early universe expansion rate relative to its current value.

Another tension where the dark matter decaying scenarios might help is regarding the amplitude of matter fluctuations $S_8$ between cosmic microwave background (CMB) and gravitational lensing,
as the value measured currently is smaller than LCDM expectations based on CMB  \citep{2018PhRvD..98d3526A,2021PhRvD.104l3533A}.
This scenario will be addressed by EUCLID studies of weak lensing \citep{2021JCAP...10..040H}.  Depending on the results, one might need to invoke new physics in the dark sector and decaying dark matter in particular \cite{Poulin:2016nat}.

Here we demonstrate that we can set constraints on both the lifetime and the characteristic kick velocity
imparted by the decay
from galactic dynamics.
The kicks can significantly deplete the dark matter in low mass subhalos and alter the subhalo mass function of Milky Way like galaxies. We measure the radial motion component of halo stars in a specified shell of matter to constrain the change in Galactic mass and constrain the dark matter lifetime.

Hierarchical structure formation within the cold dark matter paradigm is a noisy and complex process.
A galaxy with a non-zero rate of change of mass will
leave an imprint on the motions of stars within it.
Bulk radial motion can be induced directly by the complex orbits of accreting or orbiting material, or by the existence of breathing modes excited by in-falling material \citep{2014MNRAS.440.1971W}.
A population of stars that, to begin with, are in equilibrium with the Galaxy will drift radially outwards if the mass of the Galaxy decreases, or drift inwards if the mass increases. If the change of potential is slow, and the angular momentum is an adiabatic invariant during this change,
than the net average radial motion in a spherical shell is proportional to the radius $r$ of the shell and to the fractional rate of change of mass $M$ enclosed by the shell  \citep{2022RNAAS...6...26L},
\begin{eqnarray}
V_{R}\equiv \frac{{\rm d} r}{{\rm d} t} & = &  -\left(\frac{\dot{M}}{M}\right) r \label{equ:vr1}
\\
&\approx&   \left(\frac{\dot{M}/M}{{\rm Gyr^{-1}}}\right) \left(\frac{r}{{\rm kpc}}\right) \kms.
\label{equ:vr2}
\end{eqnarray}
Observationally, this can be detected by measuring the radial velocity of stars in the stellar halo. This offers the possibility to constrain the rate of change of mass in the Galaxy and the
processes associated with it, such as the  decay of dark matter.

In addition to decaying dark matter, a galaxy can gain or lose mass in a given radius for various reasons, but these are either confined to the inner galaxy or are quite small. In the hierarchical structure formation paradigm, galaxies are formed by accretion and merger events that lead to the growth of their mass with cosmic time. This sudden increase of mass can trigger inward radial motions of stars.
However, the fractional rate of change of mass is high in the first $1-2$ billion years and decreases progressively with time.
At late times, feedback from bursty star formation and supermassive
black holes can generate an outflow of gas from the central regions of the Galaxy and trigger an outward radial motion of stars \citep{2012MNRAS.421.3464P}, but this change is mostly confined to the inner regions of the Galaxy.
Galaxies lose mass over billions of years through baryonic radiative processes but the implied radial
motion is of order $\langle V_{R}\rangle \sim 0.03 \kms$ for a Milky Way-sized galaxy \citep{2022RNAAS...6...26L}.

Stellar halo stars extending up to 100 kpc and beyond \citep{2008A&ARv..15..145H} are ideal
targets for constraining the rate of change of mass in Milky Way sized galaxies. In order for this to work, the mean radial motion of stellar halo stars in the absence of change in galactic mass should be as close to zero as possible.
However it is not clear if that is true.
In the current $\Lambda$CDM paradigm of structure formation, the stellar halo is formed by accretion of
satellites onto the host galaxy \citep{2005ApJ...635..931B}. Over cosmic time, as the satellites
disrupt and phase mix, the mean radial motion of the stellar halo is expected to be close to zero.
However, not all satellites are fully phase-mixed and significant substructure can be seen in the Galaxy both in position and velocity space \citep{2008ApJ...689..936J} and this can lead to non zero mean radial motion.
The question is how large is it? To answer this, we make use of
N-body simulations and investigate the mean radial motion of stars in simulated stellar halos. We compare this with the motion expected in the scenario where dark matter undergoes
decay and discuss the physical implications of our results. Finally, we discuss the observational challenges for conducting
such a study and if the current and future observational
facilities are sufficient equipped to do so.

\section{Methods}
In this paper we study the bulk radial motion of stars in the
in the stellar halos
and for this we make use of N-body simulations and these are described in \autoref{sec:stellar_halos}.
If the stars in the stellar halo are in equilibrium then the mean radial velocity should be zero. However, some accretion events of the stellar halo are not well mixed in phase space and have not reached an equilibrium. These show up as substructures in the phase space and are associated with significant non-zero bulk motion. Since we are interested in the equilibrium component of the stellar halo, we identify and get rid of the substructures using a clustering algorithm. This is described in \autoref{sec:clustering}.
For certain halos, although the mean motion of stars in a shell is not zero, the distribution of radial velocities is quite symmetrical about the mean radial velocity. Hence, we devise an alternate
scheme to measure the central velocity of stars in a shell and this is described in
\autoref{sec:central_velocity}.

\begin{table}[htb!]
\caption{Simulated stellar halos. The halos starting with \texttt{bj\_} are from \citet{2005ApJ...635..931B} while those starting with \texttt{fire\_} are from \citet{2020ApJS..246....6S}.}
\begin{tabular}{lll}
Name & Accretion & Simulation \\
& history & type \\
\hline
\texttt{bj\_2}  & $\Lambda$CDM & Idealized \\
\texttt{bj\_5}  & $\Lambda$CDM & Idealized \\
\texttt{bj\_7}  & $\Lambda$CDM & Idealized \\
\texttt{bj\_9}  & $\Lambda$CDM & Idealized \\
\texttt{bj\_10} & $\Lambda$CDM & Idealized \\
\texttt{bj\_12} & $\Lambda$CDM & Idealized \\
\texttt{bj\_14} & $\Lambda$CDM & Idealized \\
\texttt{bj\_15} & $\Lambda$CDM & Idealized \\
\texttt{bj\_17} & $\Lambda$CDM & Idealized \\
\texttt{bj\_20} & $\Lambda$CDM & Idealized \\
\texttt{bj\_lowl}     & Artificial & Idealized \\
\texttt{bj\_highl}    & Artificial & Idealized \\
\texttt{bj\_rad}      & Artificial & Idealized \\
\texttt{bj\_circular} & Artificial & Idealized \\
\texttt{bj\_old}      & Artificial & Idealized \\
\texttt{bj\_young}    & Artificial & Idealized \\
\texttt{fire\_m12f} & $\Lambda$CDM & Cosmological \\
\texttt{fire\_m12i} & $\Lambda$CDM & Cosmological \\
\texttt{fire\_m12m} & $\Lambda$CDM & Cosmological \\
\hline
\end{tabular}
\label{tab:dataset}
\end{table}

\subsection{Simulated stellar halos}
\label{sec:stellar_halos}
To study the radial velocity of stars in the stellar halo, we make use of N-body simulations. We use three different types of simulations, and these are listed in \autoref{tab:dataset}. First, is a suite of 10 stellar halos simulated by \citet{2005ApJ...635..931B} (BJ05), named as \texttt{bj\_X}
with \texttt{X} $\in \{2,5,7,9,10,12,14,15,17,20\}$.
These have accretion histories derived from a semi-analytical scheme in accordance with the $\Lambda$CDM cosmology.
Here, a stellar halo is built up entirely by accretion of satellites. The satellites are modelled by N-body particles evolved individually in an analytical potential.
Hence, they are called idealized simulations.
Baryons are embedded deep in the inner regions. This is modelled by assigning a mass-to-light ratio to each N-body particle
based on its energy, with more tightly bound particles having lower mass-to-light ratio.
Second, is a suite of six stellar halos that were simulated by \citet{2008ApJ...689..936J} (JB08) but with artificial accretion histories, \texttt{lowl} (made up of predominantly low luminosity satellites), \texttt{highl} (made up predominantly high luminosity satellites), \texttt{old} (made up of predominantly old accretion events), \texttt{young} (made up of predominantly young accretion events), \texttt{rad} (made up of  accretion events predominantly on radial orbits), \texttt{circ} (made up of  accretion events predominantly on circular orbits). Except for the accretion history, the JB08 halos are otherwise simulated in the same way as the BJ05.
Third, we use 3 Milky Way sized galaxies simulated by the FIRE team \citep{2020ApJS..246....6S,2018MNRAS.480..800H,2016ApJ...827L..23W}, \texttt{fire\_m12f}, \texttt{fire\_m12i}, \texttt{fire\_m12m}.  These are state of the art hydrodynamical cosmological simulations including physical processes such as cooling, star formation and feedback.

\begin{figure*}
\centering \includegraphics[width=0.95\textwidth]{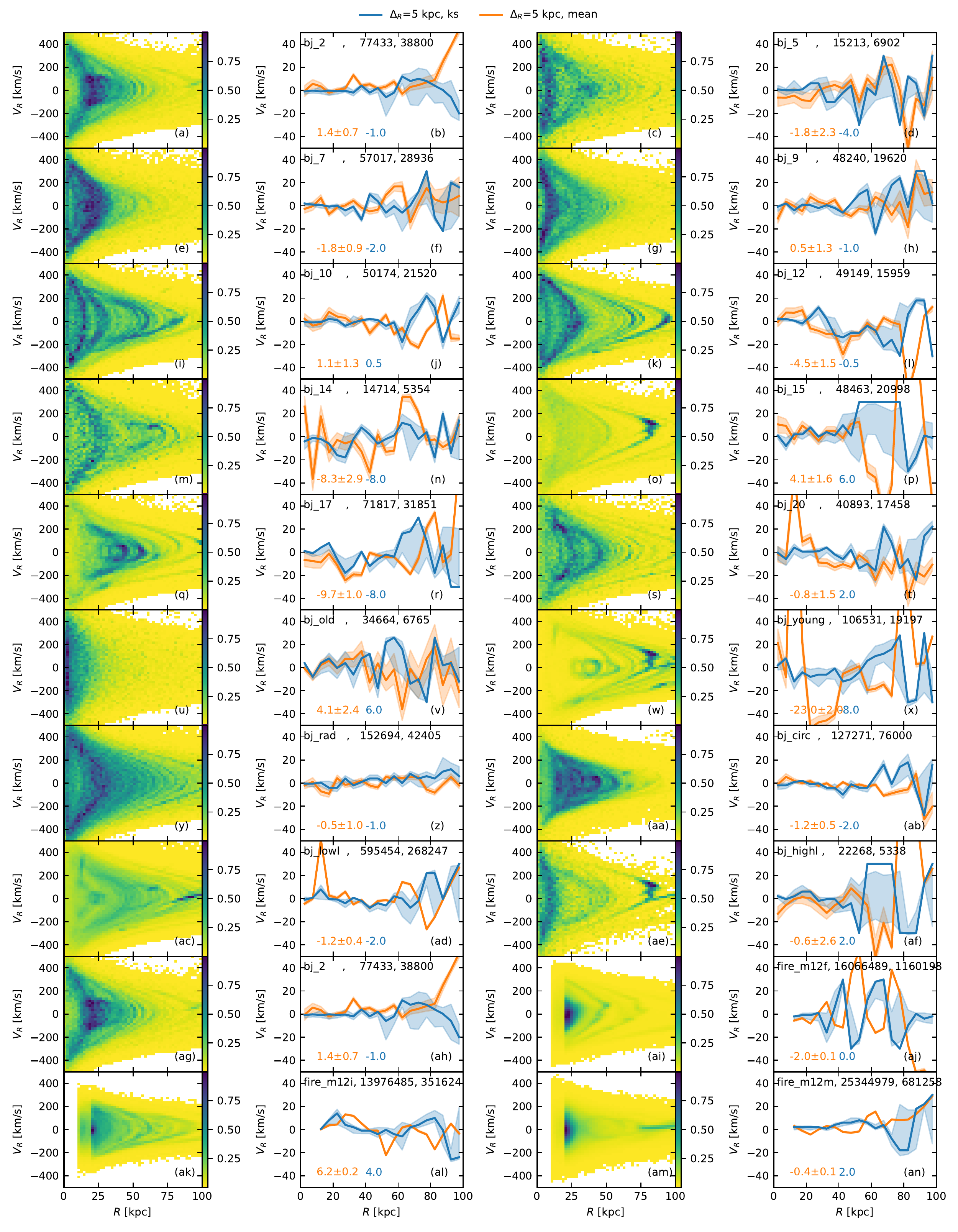}\caption{Radial velocity distribution of N-body particles in the stellar halos simulated by \citep{2005ApJ...635..931B}, \citep{2008ApJ...689..936J} and \citep{2020ApJS..246....6S}. Each N-body particle has a star forming mass  associated with them and the distributions are weighted acording to them.
First and third columns show distribution of stars in $(r,V_r)$ plane in the spherical Galactocentric coordinates. Second and fourth columns show mean radial velocity measured in spherical shells (width 5 kpc) as a function of radius (orange line).
Shown alongside (blue line) is the velocity about which the distribution is symmetric. The total number of stars
are denoted on the top, followed by number of stars in shell $15<r/{\rm kpc}<45$. For the same shell, the text at the bottom denotes, mean $V_r$, the error on the mean and  the central velocity based on symmetry.
\label{fig:r_vr}}
\end{figure*}

\begin{figure*}
\centering \includegraphics[width=0.95\textwidth]{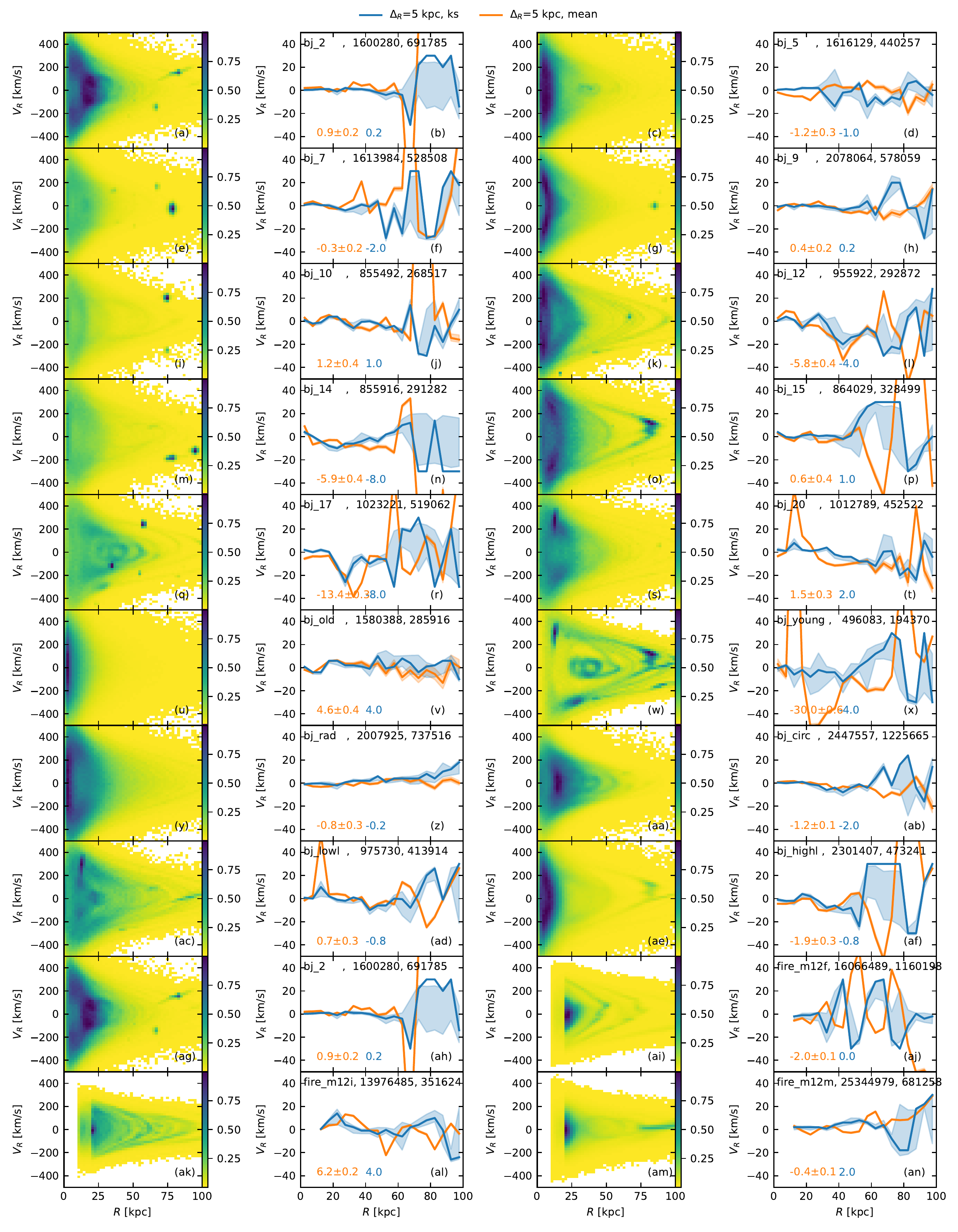}\caption{Same as \autoref{fig:r_vr} but for equal stellar mass particles. The code {\it GALAXIA} was used to spawn equal stellar mass particles from N-body particles with a given star forming mass.
\label{fig:r_vr_galaxia}}
\end{figure*}

\begin{figure*}
\centering \includegraphics[width=0.95\textwidth]{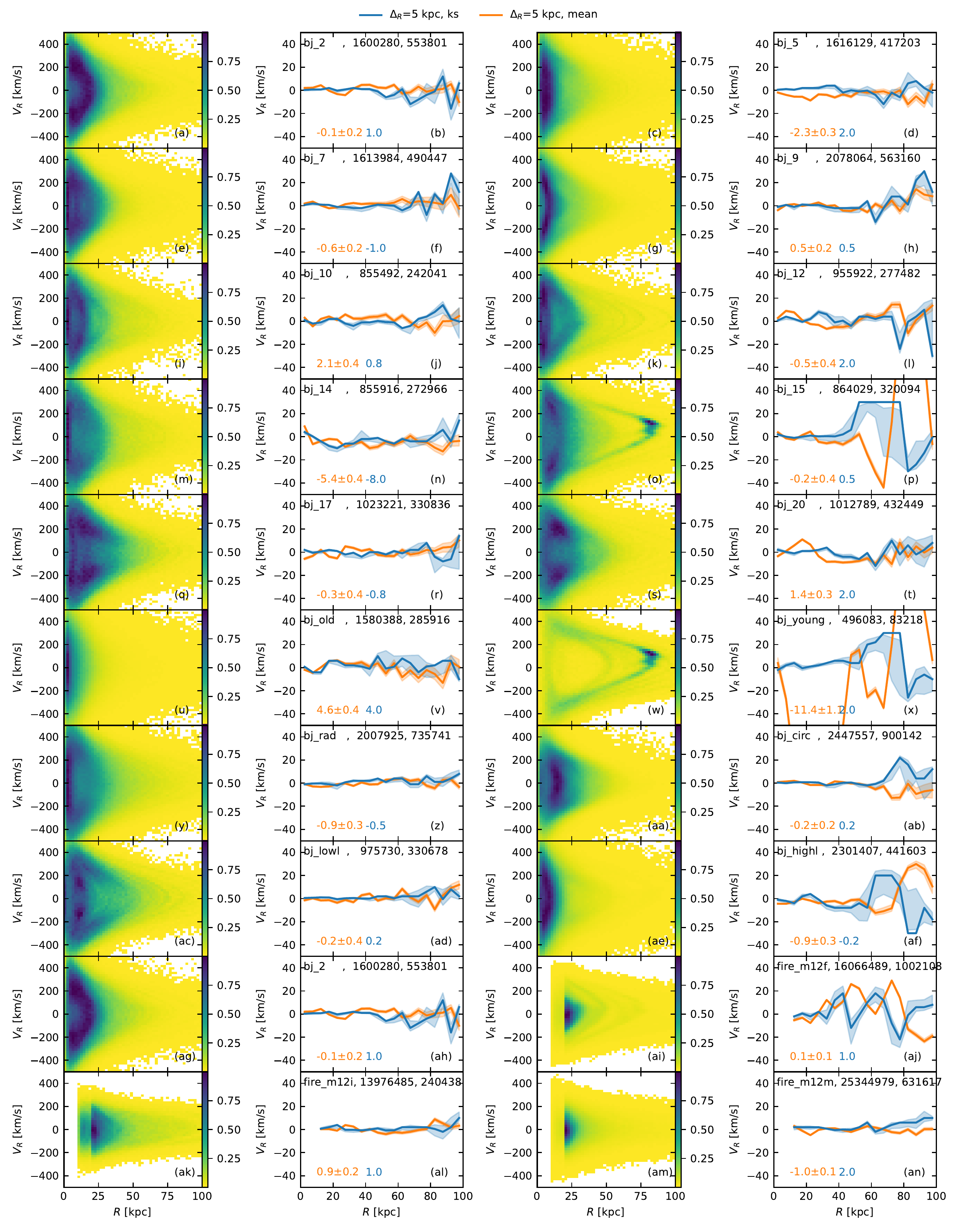}\caption{Same as \autoref{fig:r_vr_galaxia} but by filtering out substructures by using the clustering algorithm {\it ENLINK}.
\label{fig:r_vr_galaxia_cluster}}
\end{figure*}

\subsection{Clustering}
\label{sec:clustering}
To identify and remove substructures in the stellar halo,
we use the {\it ENLINK} clustering algorithm \citep{2009ApJ...703.1061S} which will be publicly available \footnote{\url{https://github.com/sanjibs/enlink}}. For examples of its application to BJ05 halos see
\citet{2010ApJ...722..750S} and \citet{2011ApJ...728..106S}.
We apply it over the six dimensional $(x,y,z,v_x,v_y,v_z)$ phase-space specified in
Cartesian coordinates. The main feature of {\it ENLINK} that is useful for our application is its ability to identify structures of arbitrary shape and size in any given multidimensional space. Unlike most other clustering algorithms that use a global metric, {\it ENLINK} makes use of a locally adaptive metric based on the idea of Shannon
entropy and calculated using a binary space partitioning tree \citep{2006MNRAS.373.1293S}.

As mentioned earlier, in the BJ05 and JB08 halos each N-body particle has different stellar mass. It is difficult to do clustering analysis on particles with unequal weights.
This is because an isolated particle of large weight will spuriously appear as a region of overdensity.
Hence, for these halos we use the code GALAXIA \citep{2011ApJ...730....3S} to sample star particles from these simulations. The number of stars spawned by an N-body particle is equal to its total stellar mass divided the mean mass of a star for a given stellar initial mass function (IMF).
The main advantage of using GALAXIA is that
it samples the stars in the six dimensional phase space, hence the sampled stars have
kinematics consistent with the original simulation.

\subsection{The central radial velocity based on symmetry}\label{sec:central_velocity}
For certain accretion events stars are not
distributed over the full available phase space of the orbit.
This means that at any given $r$ the mean motion is non zero.
However, the distribution of radial velocity is symmetrical, and the center of symmetry is close to zero. To compute the central velocity, we  divide the sample into two
about a chosen center of symmetry. Next, we minimize the
the two sample Kolmogorov-Smirnoff statistics
\be
D_{n,m}=\frac{nm}{n+m}{\rm sup} |F_{1,n}(v_r)-F_{2,m}(v_r)|
\ee
to locate the  center of symmetry. Here, $F_{1,n}$ and $F_{2,m}$ are the cumulative distribution function of the first and the second sample, and $n$ and $m$ are the respective number of data points in each of the samples.

\section{Mean radial velocity in simulated stellar halos}
We begin by studying the radial velocity distribution of simulated stellar halos. \autoref{fig:r_vr} shows the distribution of stars in the Galactocentric $(r,V_r)$ space, where $r$ is the radial distance and $v_r$ the radial velocity (panels in first and third columns).
Mean radial velocity $\langle V_r \rangle$ measured in spherical shells of width $\Delta R$ as function of radius $r$ is shown in panels of second and fourth column (orange line).
The central radial velocity, measured as the velocity about which the radial velocity distribution is symmetric, is also shown alongside (blue line).
The 16 and 84 percentile spread about the estimated mean and central velocity are denoted by the shaded region. The spread was estimated using the technique of bootstrapping.  The mean and central radial velocity
for stars in shell $15<r/{\rm kpc}<45$ is shown in the bottom right of each panel.
In \autoref{fig:r_vr}, for the idealized halos the stars are weighted by the star forming mass of each N-body particle and the bound satellites are removed.
Unlike the idealized halo the cosmological halo also has disc stars.
To get rid of disc stars, we
restrict the analysis to stars with $(R > 20\; {\rm kpc})\; {\rm or}\; (|z| > 10\; {\rm kpc})$. This is the reason for the vertical streaks at $r=10$ kpc and $r=20$ kpc in the \texttt{fire} halos.

In \autoref{fig:r_vr}, significant substructure in the $(r,V_r)$ space can be seen. The mean radial velocity is also found to show significant fluctuations.
Next, instead of the N-body particles we repeat the analysis with stellar particles of equal stellar mass  spawned by the code {\sl Galaxia}.
Results are shown in  \autoref{fig:r_vr_galaxia}.
Bound satellites were not removed and can be seen as dense knots. Substructures are not as clear as
before and this is due to two reasons. First,
the bound satellites being very dense increase the range of density being mapped by the color scale, and this lowers the contrast of the less dense substructures. Second, the star spawning process of
{\sl Galaxia} also leads to some added scatter of stars in the phase space. In spite of these minor differences, the mean radial profiles are very similar to \autoref{fig:r_vr}. Next we use the ENLINK clustering algorithm to remove the substructures and retain only the dominant smooth component of the halo.
These results are shown in \autoref{fig:r_vr_galaxia_cluster}. The distribution in $(r,V_r)$ space is much smoother and the mean radial velocity profiles have markedly smaller fluctuations.

\begin{figure}[!htb]
\centering \includegraphics[width=0.49\textwidth]{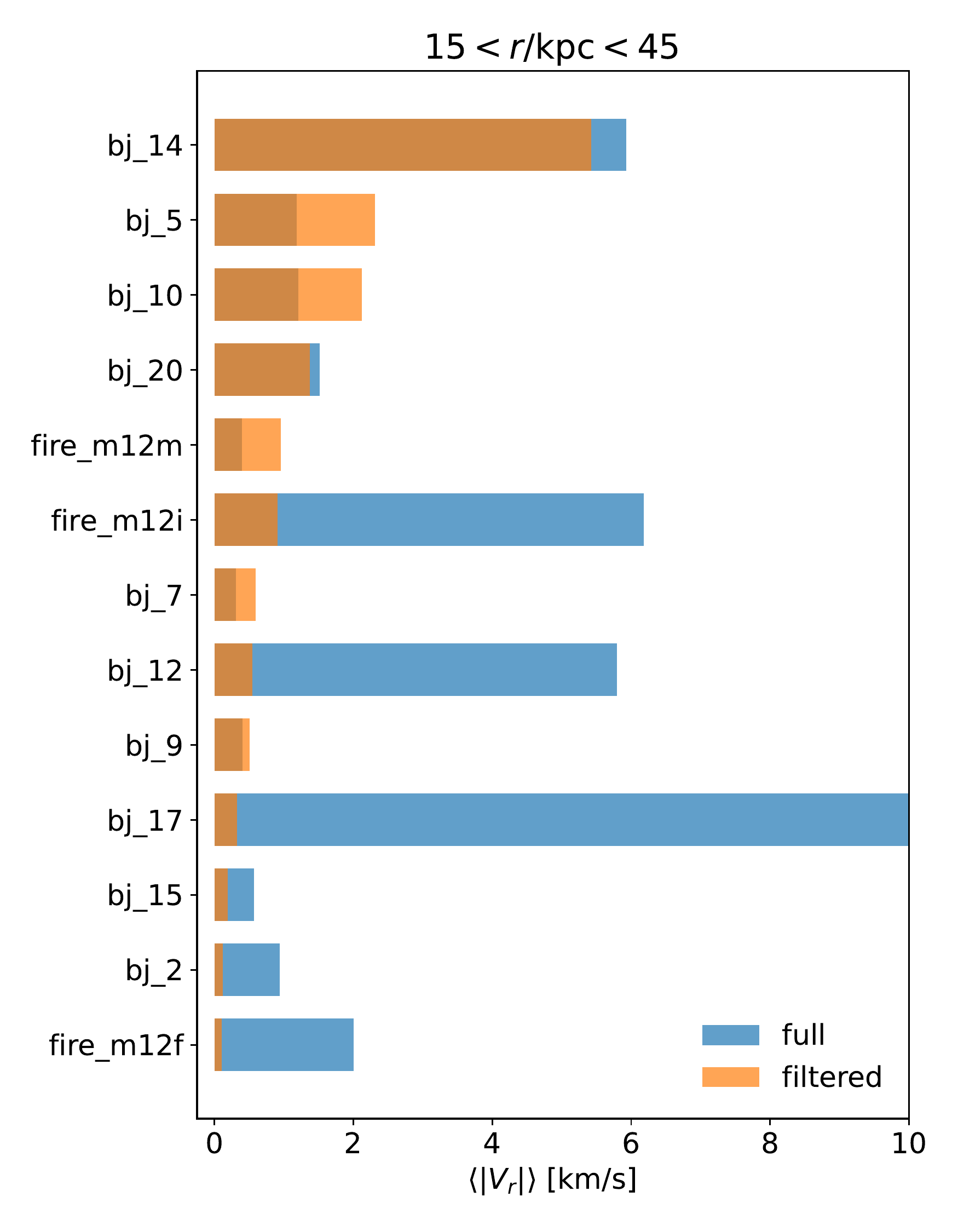}\caption{Absolute mean radial velocity of stars in a spherical shell for different simulated $\Lambda$CDM stellar halos.
Results of the full sample are compared with the
sample where substructures were filtered out.
\label{fig:stats1}}
\end{figure}

\begin{figure}[!htb]
\centering \includegraphics[width=0.49\textwidth]{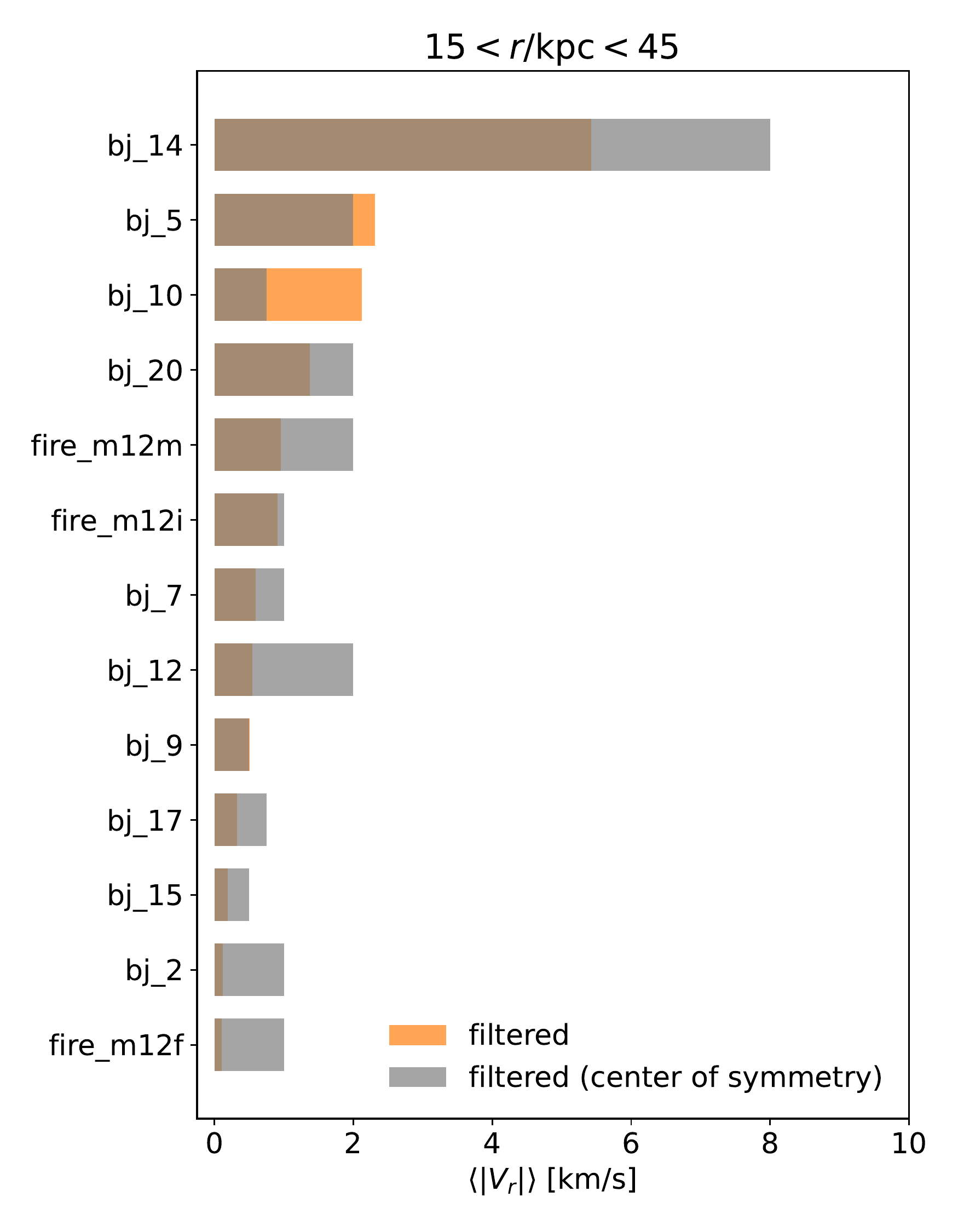}\caption{Absolute central radial velocity (see \autoref{sec:central_velocity}) of stars in a spherical shell for different simulated $\Lambda$CDM stellar halos.
Results of the full sample are compared with the
sample where substructures were filtered out.
\label{fig:stats2}}
\end{figure}

\subsection{Mean radial velocity in shell $15 < r/{\rm kpc} < 45$.}\label{sec:mean_radial_motion}
We now focus on the mean radial velocity in
the spherical shell $15 < r/{\rm kpc} < 45$ centered around $r=30$ kpc.
We choose this radius for the following reasons.
From \autoref{equ:vr1} it is clear that the mean radial velocity due to decay
is proportional to radius. However, the number density of stars in the Galaxy decreases
sharply with radius, making it difficult to find a large number of stars to observe.
The density of stars in the stellar halo is well approximated by Hernquist profile \citep{2005ApJ...635..931B},
it is high in the center and decreases with radius (varying as $r^{-1}$ at small $r$ and $r^{-4}$ at large $r$).
The further the stars are, the more exposure time is required to observe them.
Additionally, the stellar halo is also less phase mixed at large $r$,
which is due to the relaxation time of stars there being large.
This can be seen \autoref{fig:r_vr_galaxia} and \autoref{fig:r_vr_galaxia_cluster}.
Finally, for $r<15$ kpc, the stellar population is dominated by disc stars, which can have
significant bulk motion due to non axis-symmetric structures like the spiral arms
and the bar. Velocity fluctuations in the disc of the order of 5 to 10 km s$^{-1}$ were shown by
\citep{2018A&A...616A..11G, 2019MNRAS.482.4215K, 2019MNRAS.489.4962K}.
Also the orbiting satellites, like the Sagittarius dwarf galaxy  can disturb the disc, an example of this
is the $(z,V_z)$ phase space spiral \citep{2018Natur.561..360A,2018MNRAS.481..286L, 2019MNRAS.486.1167B, 2019MNRAS.485.3134L}.

In \autoref{fig:stats1}, we show the shell radial speed using a bar blot for different stellar halos simulated with $\Lambda$CDM accretion history.
Results for both the full sample and the sample where substructures were filtered out are shown together.
Note, we analyse speed instead of velocity. This is  to improve the statistics as we only have 13 halos.
We assume that mean radial velocity of stars in a shell is equally likely to be either positive or negative. This is very close to true for our
sample where the mean velocity of shells was found to be close to zero.
It can be seen that for the full sample the shell speed is typically small (median over 13 halos being 1.2 km s$^{-1}$), but for four halos it is larger  than 4 km s$^{-1}$. After filtering out substructures significant reduction in the speed can be seen, median shell speed being 0.6 km s$^{-1}$ and
only one halo having speed above 4 km s$^{-1}$.
This implies that 75\% of halos have $\langle V_r \rangle < 0.6$ km s$^{-1}$.
In \autoref{fig:stats2} we compare the mean shell radial speed with the central velocity based on symmetry. Although the mean and central radial velocity values differ slightly from halo to halo, but overall the two values are very similar for most halos.

\begin{figure}[!htb]
\centering \includegraphics[width=0.49\textwidth]{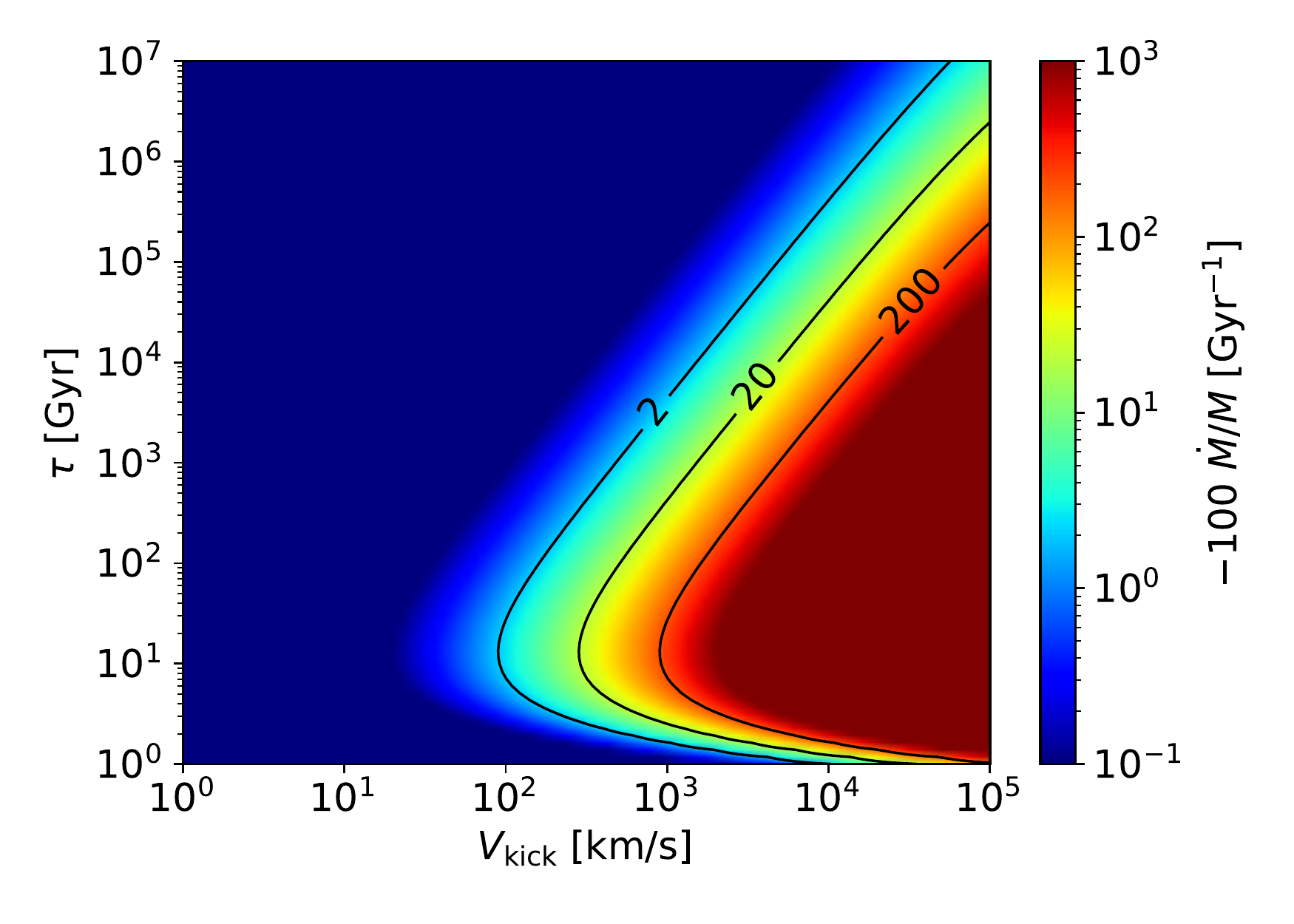}\caption{Mass loss rate due to decay of dark matter for a dark matter halo as a function of $V_{\rm kick}$ and decay lifetime $\tau$.
Mass loss rate is shown for a sphere of radius 30 kpc and 13 Gyr after the formation of the halo. The solid lines are contours for mass loss rates per Gyr of 2\%, 20\% and 200\%. The result is for an NFW halo with a virial mass of $0.8 \times 10^{12}$  and concentration parameter $c=20$, but approximated by a Plummer model following \citet{2008MNRAS.388.1869A}.\label{fig:DDM_constrain}}
\end{figure}

\section{Implications for detecting dark matter decay}
Dark matter invisible decay is currently unconstrained by dark matter detection experiments both direct and indirect. Here we consider two decay mechanisms and explore if we can detect them using kinematics of stars in the stellar halo. First mechanism is the full decay of a dark matter particle into some form of radiation ($BR = 1$). Second mechanism is the  decay into some radiation and a daughter particle lighter than the dark matter.
An example of the former scenario is the 2-body decay of the supersymmetric scalar partner of the axion into two axions, where the axions can be dark radiation \citep{Kawasaki:2007mk}. Examples of the latter scenario include models where the dark matter is coupled to a dark photon/dark Z' \citep{Boehm:2003hm} or, in the case of supersymmetry, sneutrino decaying into a pair neutralino-neutrino or a pair gravitino-neutrino \citep{Kim:2022gpl}. These different channels may eventually lead to visible signatures, including in ICECUBE if the dark matter produces high energy neutrinos, but they may also stay invisible for some parts of the parameter space \footnote{We disregard $\tilde{G} \rightarrow \chi + \gamma$ as this could be in principle constrained by traditional means.}.

In both of the above scenarios, the mass enclosed by a shell of any given radius will decrease with time.
In the first scenario there is a direct decrease of mass
enclosed by a shell. In the second scenario,  the decay imparts a kick to the daughter particle, which induces an expansion of the dark matter halo.
In principle, the change of mass can be detected
as a non-zero outward radial motion of stars.
In \autoref{sec:mean_radial_motion}, we saw that the median
expected radial shell speed of stellar halo at a radius of 30 kpc is 0.6 km s$^{-1}$. Using \autoref{equ:vr1}, this translates to a  mass loss rate $\dot{M}/M$ of 0.02 per Gyr.
Hence, if the mass loss rate due to decay is higher than 2\% per Gyr then it should be detectable using mean motion of stellar halo stars.

The dark matter decay is characterized by the lifetime of decay $\tau$ and, in case where there is a daughter particle, the kick velocity $V_{\rm kick}$ imparted to the daughter particle. We now explore in detail the region of the parameter space over which dark matter decay should be detectable using the mean motion of stellar halo stars.
In general, for dark matter decaying with lifetime $\tau$
the number of unstable dark matter particles $N$ at a time $t$
since the formation of the halo is given by
\begin{equation}
N=N_0 \exp(-t/\tau).
\end{equation}
and the rate of change by ${\rm d}N/{\rm d}t=-N/\tau$.
Here, $N_0$ the initial number of unstable dark matter
particles at $t=0$.
For dark matter decaying purely into radiation we have
$-\dot{M}/M=1/\tau$. A limit of $\dot{M}/M>0.02\ {\rm Gyr}^{-1}$ implies $\tau<50$ Gyr.

For the case where dark matter decays into a daughter particle, following previous studies \citep{2008MNRAS.388.1869A,2022arXiv220111740M},
we assume a dark matter particle $\chi$ of mass $m$ decays with lifetime $\tau$ into a massive daughter particle $\chi'$ of mass $m'$ and a lighter probably massless dark radiation species $\gamma'$,
\begin{equation}
\chi \rightarrow \chi' + \gamma'.
\end{equation}
Due to conservation of momentum, the decay imparts a velocity kick of
\begin{equation}
V_{\rm kick}=\epsilon c,
\end{equation}
where $\epsilon=(m-m')/m$ is the mass splitting factor.
The kick increases the velocity dispersion of dark matter particles, which in turn will force the halo to expand. Given the dynamical time is in general smaller than the decay lifetime, the halo should quickly virialize such that the expansion can be considered to be adiabatic.

Approximating the dark matter halo with a Plummer model,
\citet{2008MNRAS.388.1869A} derived the increase of its scale radius $r_p$ with time as
\begin{eqnarray}
\frac{{\rm d} r_p}{{\rm d}t} &=&\frac{64 r_p^2 c^2}{3\pi GM^2}\frac{\exp[-(t+t_f)/\tau]}{\tau}\times \nonumber \\
&&\left[\frac{\chi}{1+\chi}-\left(1+\frac{3\pi GM}{64c^2 r_p}\right)\frac{\chi(2+\chi)}{2(1+\chi)}\right].
\end{eqnarray}
For Plummer model, the mass enclosed by a radial shell is given by
\begin{equation}
M(r)=M_{\rm vir} \frac{r^3}{(r_p^2+r^2)^{3/2}}
\end{equation}
Due to expansion of the dark matter halo, the mass enclosed in a given radial shell should decrease. We estimate this
taking the derivative of $M(r)$ with time, which gives
\begin{equation}
\frac{\dot{M}}{M}=\frac{{\rm d} r_p}{{\rm d}t}\frac{3r_p}{r_p^2+r^2}.
\end{equation}
In \autoref{fig:DDM_constrain}, we explore the mass loss rate at radius of 30 kpc for
a Milky Way mass halo
as a function of parameters $\tau$ and $V_{\rm kick}$. Following \citet{2008MNRAS.388.1869A}
we approximate an NFW halo with a Plummer model, for an NFW halo with
scale radius $r_s$ and concentration parameter $c$ the equivalent Plummer scale radius $r_p$
is given by
\begin{equation}
r_p=r_s\frac{3\pi}{16}
\left\{\frac{[\ln(1+c)-c/(1+c)]^2}{1-1/(1+c)^2-2\ln(1+c)/(1+c)}\right\}.
\end{equation}
\autoref{fig:DDM_constrain} shows
that for $V_{\rm kick}<100$ km s$^{-1}$ the mass loss rate per Gyr is less than 2\%. However, for $V_{\rm kick}>100$ km s$^{-1}$ the rate increases steadily with $V_{\rm kick}$ for any given $\tau$. For a given $V_{\rm kick}$ the mass loss rate seems to be maximum for $\tau$ close to 10 Gyr.
Contour lines for mass loss rate of 2\%, 20\% and 200\% are shown in the figure.
The region of the parameter space over which
dark matter decay should be detectable, that is mass loss rate is greater than 2\%, can be seen from \autoref{fig:DDM_constrain}, it is right of the line labelled 2.
For a mass loss rate as small as 2\% per Gyr, we can rule out for $\tau=10$ Gyr,
$V_{\rm kick} > 100$ km\ s$^{-1}$.
In contrast, a $V_{\rm kick} \approx 10^4 {\rm km\ s}^{-1}$ is required to resolve the $H_0$ tension \citep{2019PhRvD..99l1302V} while
a $V_{\rm kick} \approx 10^5 {\rm km\ s}^{-1}$ is required to resolve the $S_8$ tension \citep{2021PhRvD.104l3533A}.
Hence, we are sensitive to values of
$V_{\rm kick}$ that are much lower than that required to resolve the tensions of Hubble parameter $H_0$ and the amplitude parameter S8.

Using the observed population of Milky Way satellites \citet{2022arXiv220111740M} placed constraints of $\tau < 18$ Gyr (29 Gyr) for $V_{\rm kick}=20$ km\ s$^{-1}$ (40  km\ s$^{-1}$). This is stricter than the limits that we can set based Milky Way's stellar halo kinematics. This is because the effect of a given kick is stronger for smaller subhalos due to their shallower potential wells.
However,  significant assumptions related to poorly understood baryonic processes are needed in order to connect the subhalos in simulations to luminous satellite galaxies.
In this sense  our results based on an independent
physics are useful and play a complementary role.

\begin{figure}[!htb]
\centering\includegraphics[width=0.49\textwidth]{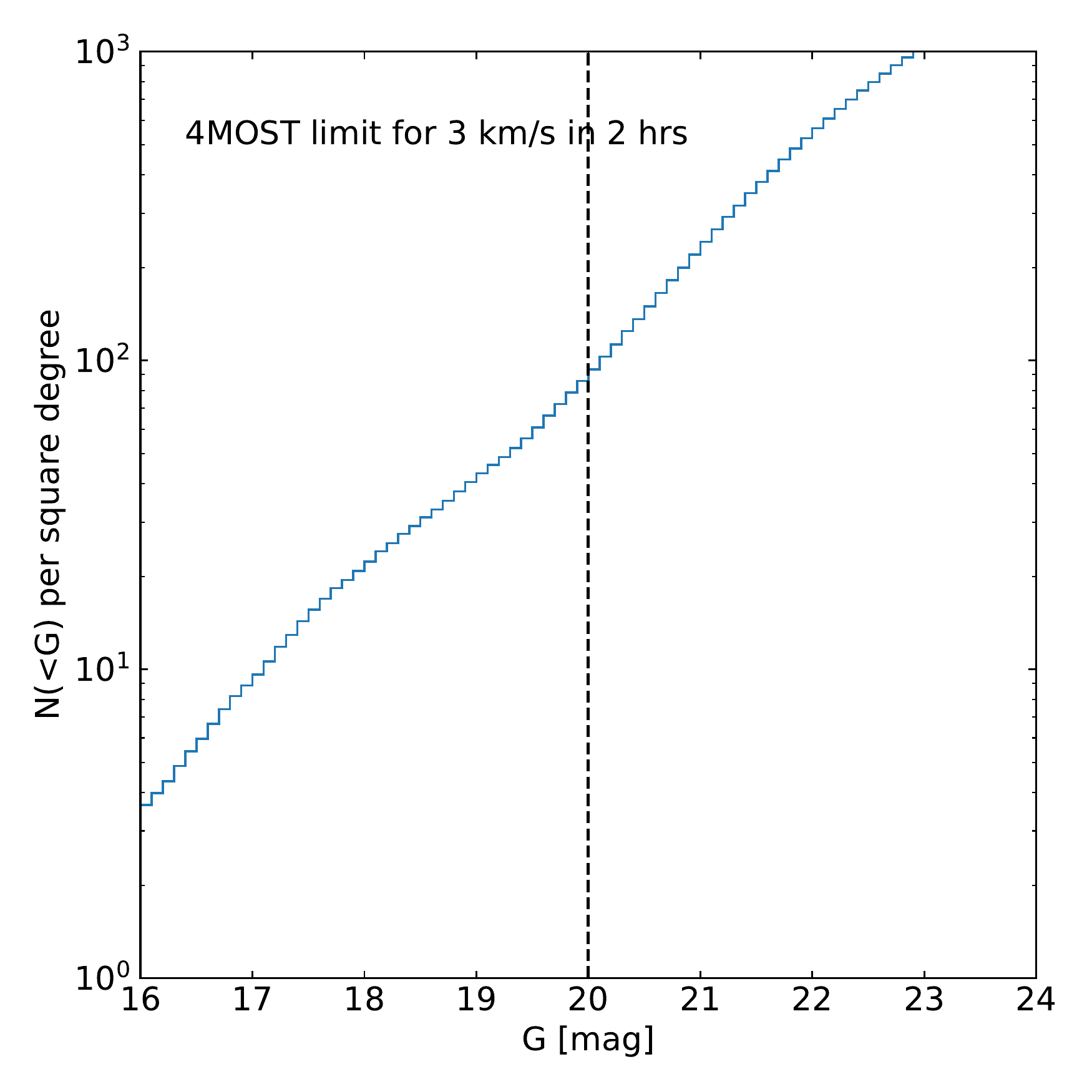}\caption{Cumulative number of stars lying in the shell with $(15<R/{\rm kpc}<45)$ and having $(|z|/{\rm kpc}>10)$, as a function of $G$ band apparent magnitude based on a simulation of the Milky Way by the code {\it GALAXIA}.
\label{fig:stellar_halo_cumulative}}
\end{figure}

\section{Observational feasibility}
We now look into the feasibility of conducting a study to measure the
mean radial motion of Milky Way halo stars in the $15<r/{\kpc}<45$ spherical shell.
Two independent arguments suggest that of the order of a million stars would be required to detect mean radial motion of  greater than equal to 0.6 km s$^{-1}$.
First, a large sample of halo stars is required to do clustering and filter out substructures, without which the radial motion would be too noisy. \autoref{fig:r_vr_galaxia_cluster} shows that of the order of 1 million stars is sufficient to suppress the noise due to substructures.
Secondly, to measure a mean motion of 0.6 km s$^{-1}$, uncertainty of less than 0.1 km s$^{-1}$ is desirable. To achieve this, given that the radial velocity dispersion of stars in the halo is 140 km s$^{-1}$ \citep{2003A&A...409..523R},
of the order of 1 million stars are required.

\autoref{fig:stellar_halo_cumulative}
shows the cumulative number density of
stars lying in shell $15<R/{\rm kpc}<45$
as a function of $G$ band magnitude based on simulation of the Milky Way by {\it GALAXIA}.
There are close to 90 stars per square degree for $G<20$.
A multi-object spectroscopic survey in either north or southern hemisphere targeting 10,000 to 15,000 square degrees can easily observe close to a million stars.

We now discuss the exposure time for
each pointing and the total duration required to complete a million star survey of halo stars. Given the intrinsic radial velocity dispersion of halo stars is close to 140 km s$^{-1}$, the requirements on the precision of radial velocity measurements of individual stars are less stringent. Even a precision of 10-20 km s$^{-1}$ should be sufficient.
Several wide-field surveys have measured stellar radial velocities
for millions of stars; these include APOGEE \citep[0.6M, ][]{2017AJ....154...94M}, GALAH+ \citep[0.6M,][]{2021MNRAS.506..150B}, LAMOST \citep[7M,][]{2012RAA....12..723Z} and {\it Gaia} \citep[33M][]{2021A&A...649A...1G}. All of these surveys have been carried out on moderately small telescopes (1-4m diameter) or on mediocre sites, or both, and thus the magnitude limit ($V\lesssim 15$) is too bright for the proposed experiment, yielding a typical measurement accuracy of $1-10$ km s$^{-1}$ depending on the survey. With the dawn of wide-field positioners on 8m class telescopes (e.g. PFS on Subaru 8m, WST 12m in Chile), or 4m class telescopes on exceptional sites, e.g. 4MOST on VISTA \citep{2019Msngr.175....3D}, we are entering a new era where accurate stellar radial velocities will be routinely accessible down to fainter magnitude limits.

We focus on 4MOST for a more detailed  study to demonstrate the feasibility of our experiment. This is the next major ESO VLT project, to be delivered by 2025, involving a dedicated optical 4m telescope and multi-object spectrographs. 4MOST can observe 1462 stars at low spectroscopic resolution ($R = \lambda/\delta \lambda \approx 4000 - 7500$)
and 812 stars in high resolution ($R \approx 18000 - 21000$) mode.
The expected 4MOST limit\footnote{
\url{https://www.4most.eu/cms/facility/overview
}} for a 2 hour exposure in low resolution is 1 km s$^{-1}$ (1$\sigma$) at $V\sim 18$ increasing to 3 km s$^{-1}$ at $V\sim 20$, which are feasible with proper consideration of which spectral features are not affected by stellar winds \citep{2018MNRAS.481..645Z}.
In fact, the 4MOST low-resolution ($R \sim$ 4000 - 7500) halo survey \citep{2019Msngr.175...23H} is planning to
observe almost all halo giants with $G<20$ mag over 10,000 square degrees, which is about 1.5 million halo stars. Based on GALAXIA about one third of these stars (0.5 million) will be in our desired radial shell.

We now estimate the time required for 4MOST if it were to exclusively focus on
halo stars that lie in our desired shell.
4MOST has a field of view of 2.5 square degrees, and there are about 225 targets per 4MOST pointing that lie in our desired shell, which 4MOST can easily do given its high multiplexing.
With 2 hours of exposures required for a 3 km s$^{-1}$ radial velocity precision, 4MOST can acquire 4 fields per night or about 14,600 square degrees (1.3 million $G<20$ stars in our required shell) in 5 years (assuming 80\% of the time available for observations). Given the high multiplexing of 4MOST, one can in principle go fainter to say $G\approx 22$ and try to fill up all 1462 low resolution fibers with stars in our desired shell. However, in this case it is not useful to do so. The 2.5 times increase in exposure time per magnitude cancels the gains due to the increase in target density.

With a view to proposed facilities in the next decade
, e.g. the Wide-field Survey Telescope \citep[WST,][]{2017arXiv170101976E},
we note that 4MOST is mounted on the VISTA 4m telescope. For the same field of view and fibre density, a 12m class telescope as proposed for the ESO WST can do the above survey about 9 times faster.
This remarkable prospect will allow for experiments on external galaxies and many other sophisticated experiments of dark matter properties.
With large data sets of halo stars one can also learn about the aspherical nature of the halo as has been shown in simulations, e.g., twisting and stretching of halos  \citep{2021ApJ...913...36E}
or the LMC-induced sloshing of the halo \citep{2021MNRAS.506.2677E}.

\section{Conclusions}
Under the assumption that the stars in the stellar halo are in equilibrium with
potential of the Milky Way, there should be no net radial motion of stars.
Any change in mass of the Galaxy is predicted to generate bulk radial motion of stars
in the galaxy. Hence, a measurement of non zero bulk radial motion puts constraints on the rate of change of mass in the Galaxy. With this in mind, we have studied the expected bulk radial motion of stars in the stellar halo formed in accordance with the currently favoured $\Lambda$CDM model of structure formation. Our main result is that the median radial velocity
for 75\%
$\Lambda$CDM halos measured in a shell of radius $15<R/{\rm kpc}<45$ is less than 0.6 km s$^{-1}$. This implies that using stellar halo stars we can measure the
rate of change of mass provided it is greater that 2\%  per giga year. If such rate of change of mass is due decay of dark matter purely into radiation then our results suggest that we can detect decay with lifetime of
less than 50 Gyr. If the change in mass is due to the decay of dark matter into radiation and daughter particles,
then our results suggest
that we can detect a decay with kick velocity of the order of 100 km/s and a lifetime of 10 Gyr.
If kick velocity is larger than 100 km/s then one can detect decay for a wide range of lifetimes.
In order to conduct such an experiment and measure a
signal in radial motion of 0.6 km s$^{-1}$, of the order of 1 million halo stars would be required. This is feasible with the current generation of astronomical facilities like the 4m class 4MOST facility operating over a period of 5 years. Future facilities with a larger telescope aperture can do this even faster.

\section{Data availability}\label{sec:data_avail}
The code GALAXIA used for generating mock observational surveys is available at \url{http://galaxia.sourceforge.net/}.
Links to the stellar halos simulated by BJ05 and BJ07 are
also provided there.
The galaxies simulated by the fire team are available
at \url{https://fire.northwestern.edu/ananke/}.
The code ENLINK used for clustering will be available
at \url{https://github.com/sanjibs/enlink}.

\section*{Acknowledgements}
SS is funded by a Senior Fellowship (University of Sydney), an ARC Centre of Excellence for All Sky Astrophysics in 3 Dimensions (ASTRO-3D) Research Fellowship and JBH's Laureate Fellowship from the Australian Research Council (ARC). JBH is supported by an ARC Australian Laureate Fellowship (FL140100278) and ASTRO-3D.

\bibliographystyle{aasjournal}

\begin{thebibliography}{}
\expandafter\ifx\csname natexlab\endcsname\relax\def\natexlab#1{#1}\fi
\providecommand{\url}[1]{\href{#1}{#1}}
\providecommand{\dodoi}[1]{doi:~\href{http://doi.org/#1}{\nolinkurl{#1}}}
\providecommand{\doeprint}[1]{\href{http://ascl.net/#1}{\nolinkurl{http://ascl.net/#1}}}
\providecommand{\doarXiv}[1]{\href{https://arxiv.org/abs/#1}{\nolinkurl{https://arxiv.org/abs/#1}}}

\bibitem[{{Abbott} {et~al.}(2018){Abbott}, {Abdalla}, {Alarcon}, {Aleksi{\'c}},
  {Allam}, {Allen}, {Amara}, {Annis}, {Asorey}, {Avila}, {Bacon}, {Balbinot},
  {Banerji}, {Banik}, {Barkhouse}, {Baumer}, {Baxter}, {Bechtol}, {Becker},
  {Benoit-L{\'e}vy}, {Benson}, {Bernstein}, {Bertin}, {Blazek}, {Bridle},
  {Brooks}, {Brout}, {Buckley-Geer}, {Burke}, {Busha}, {Campos}, {Capozzi},
  {Carnero Rosell}, {Carrasco Kind}, {Carretero}, {Castander}, {Cawthon},
  {Chang}, {Chen}, {Childress}, {Choi}, {Conselice}, {Crittenden}, {Crocce},
  {Cunha}, {D'Andrea}, {da Costa}, {Das}, {Davis}, {Davis}, {De Vicente},
  {DePoy}, {DeRose}, {Desai}, {Diehl}, {Dietrich}, {Dodelson}, {Doel},
  {Drlica-Wagner}, {Eifler}, {Elliott}, {Elsner}, {Elvin-Poole}, {Estrada},
  {Evrard}, {Fang}, {Fernandez}, {Fert{\'e}}, {Finley}, {Flaugher}, {Fosalba},
  {Friedrich}, {Frieman}, {Garc{\'\i}a-Bellido}, {Garcia-Fernandez}, {Gatti},
  {Gaztanaga}, {Gerdes}, {Giannantonio}, {Gill}, {Glazebrook}, {Goldstein},
  {Gruen}, {Gruendl}, {Gschwend}, {Gutierrez}, {Hamilton}, {Hartley}, {Hinton},
  {Honscheid}, {Hoyle}, {Huterer}, {Jain}, {James}, {Jarvis}, {Jeltema},
  {Johnson}, {Johnson}, {Kacprzak}, {Kent}, {Kim}, {King}, {Kirk}, {Kokron},
  {Kovacs}, {Krause}, {Krawiec}, {Kremin}, {Kuehn}, {Kuhlmann}, {Kuropatkin},
  {Lacasa}, {Lahav}, {Li}, {Liddle}, {Lidman}, {Lima}, {Lin}, {MacCrann},
  {Maia}, {Makler}, {Manera}, {March}, {Marshall}, {Martini}, {McMahon},
  {Melchior}, {Menanteau}, {Miquel}, {Miranda}, {Mudd}, {Muir}, {M{\"o}ller},
  {Neilsen}, {Nichol}, {Nord}, {Nugent}, {Ogando}, {Palmese}, {Peacock},
  {Peiris}, {Peoples}, {Percival}, {Petravick}, {Plazas}, {Porredon}, {Prat},
  {Pujol}, {Rau}, {Refregier}, {Ricker}, {Roe}, {Rollins}, {Romer}, {Roodman},
  {Rosenfeld}, {Ross}, {Rozo}, {Rykoff}, {Sako}, {Salvador}, {Samuroff},
  {S{\'a}nchez}, {Sanchez}, {Santiago}, {Scarpine}, {Schindler}, {Scolnic},
  {Secco}, {Serrano}, {Sevilla-Noarbe}, {Sheldon}, {Smith}, {Smith}, {Smith},
  {Soares-Santos}, {Sobreira}, {Suchyta}, {Tarle}, {Thomas}, {Troxel},
  {Tucker}, {Tucker}, {Uddin}, {Varga}, {Vielzeuf}, {Vikram}, {Vivas},
  {Walker}, {Wang}, {Wechsler}, {Weller}, {Wester}, {Wolf}, {Yanny}, {Yuan},
  {Zenteno}, {Zhang}, {Zhang}, {Zuntz}, \& {Dark Energy Survey
  Collaboration}}]{2018PhRvD..98d3526A}
{Abbott}, T.~M.~C., {Abdalla}, F.~B., {Alarcon}, A., {et~al.} 2018, \prd, 98,
  043526, \dodoi{10.1103/PhysRevD.98.043526}

\bibitem[{{Abdelqader} \& {Melia}(2008)}]{2008MNRAS.388.1869A}
{Abdelqader}, M., \& {Melia}, F. 2008, \mnras, 388, 1869,
  \dodoi{10.1111/j.1365-2966.2008.13530.x}

\bibitem[{{Abell{\'a}n} {et~al.}(2021){Abell{\'a}n}, {Murgia}, \&
  {Poulin}}]{2021PhRvD.104l3533A}
{Abell{\'a}n}, G.~F., {Murgia}, R., \& {Poulin}, V. 2021, \prd, 104, 123533,
  \dodoi{10.1103/PhysRevD.104.123533}

\bibitem[{{Alvi} {et~al.}(2022){Alvi}, {Brinckmann}, {Gerbino}, {Lattanzi}, \&
  {Pagano}}]{2022arXiv220505636A}
{Alvi}, S., {Brinckmann}, T., {Gerbino}, M., {Lattanzi}, M., \& {Pagano}, L.
  2022, arXiv e-prints, arXiv:2205.05636.
\newblock \doarXiv{2205.05636}

\bibitem[{{Anchordoqui} {et~al.}(2022){Anchordoqui}, {Barger}, {Marfatia}, \&
  {Soriano}}]{2022PhRvD.105j3512A}
{Anchordoqui}, L.~A., {Barger}, V., {Marfatia}, D., \& {Soriano}, J.~F. 2022,
  \prd, 105, 103512, \dodoi{10.1103/PhysRevD.105.103512}

\bibitem[{{Antoja} {et~al.}(2018){Antoja}, {Helmi}, {Romero-G{\'o}mez}, {Katz},
  {Babusiaux}, {Drimmel}, {Evans}, {Figueras}, {Poggio}, {Reyl{\'e}}, {Robin},
  {Seabroke}, \& {Soubiran}}]{2018Natur.561..360A}
{Antoja}, T., {Helmi}, A., {Romero-G{\'o}mez}, M., {et~al.} 2018, \nat, 561,
  360, \dodoi{10.1038/s41586-018-0510-7}

\bibitem[{{Bland-Hawthorn} {et~al.}(2019){Bland-Hawthorn}, {Sharma},
  {Tepper-Garcia}, {Binney}, {Freeman}, {Hayden}, {Kos}, {De Silva}, {Ellis},
  {Lewis}, {Asplund}, {Buder}, {Casey}, {D'Orazi}, {Duong}, {Khanna}, {Lin},
  {Lind}, {Martell}, {Ness}, {Simpson}, {Zucker}, {Zwitter}, {Kafle},
  {Quillen}, {Ting}, \& {Wyse}}]{2019MNRAS.486.1167B}
{Bland-Hawthorn}, J., {Sharma}, S., {Tepper-Garcia}, T., {et~al.} 2019, \mnras,
  486, 1167, \dodoi{10.1093/mnras/stz217}

\bibitem[{Boehm \& Fayet(2004)}]{Boehm:2003hm}
Boehm, C., \& Fayet, P. 2004, Nucl. Phys. B, 683, 219,
  \dodoi{10.1016/j.nuclphysb.2004.01.015}

\bibitem[{Boyarsky {et~al.}(2015)Boyarsky, Franse, Iakubovskyi, \&
  Ruchayskiy}]{Boyarsky:2014ska}
Boyarsky, A., Franse, J., Iakubovskyi, D., \& Ruchayskiy, O. 2015, Phys. Rev.
  Lett., 115, 161301, \dodoi{10.1103/PhysRevLett.115.161301}

\bibitem[{{Buder} {et~al.}(2021){Buder}, {Sharma}, {Kos}, {Amarsi},
  {Nordlander}, {Lind}, {Martell}, {Asplund}, {Bland-Hawthorn}, {Casey}, {de
  Silva}, {D'Orazi}, {Freeman}, {Hayden}, {Lewis}, {Lin}, {Schlesinger},
  {Simpson}, {Stello}, {Zucker}, {Zwitter}, {Beeson}, {Buck}, {Casagrande},
  {Clark}, {{\v{C}}otar}, {da Costa}, {de Grijs}, {Feuillet}, {Horner},
  {Kafle}, {Khanna}, {Kobayashi}, {Liu}, {Montet}, {Nandakumar}, {Nataf},
  {Ness}, {Spina}, {Tepper-Garc{\'\i}a}, {Ting}, {Traven},
  {Vogrin{\v{c}}i{\v{c}}}, {Wittenmyer}, {Wyse}, {{\v{Z}}erjal}, \& {Galah
  Collaboration}}]{2021MNRAS.506..150B}
{Buder}, S., {Sharma}, S., {Kos}, J., {et~al.} 2021, \mnras, 506, 150,
  \dodoi{10.1093/mnras/stab1242}

\bibitem[{{Bullock} \& {Johnston}(2005)}]{2005ApJ...635..931B}
{Bullock}, J.~S., \& {Johnston}, K.~V. 2005, \apj, 635, 931,
  \dodoi{10.1086/497422}

\bibitem[{Cabibbo {et~al.}(1981)Cabibbo, Farrar, \& Maiani}]{Cabibbo:1981er}
Cabibbo, N., Farrar, G.~R., \& Maiani, L. 1981, Phys. Lett. B, 105, 155,
  \dodoi{10.1016/0370-2693(81)91010-8}

\bibitem[{{Carney} {et~al.}(2022){Carney}, {Raj}, {Bai}, {Berger}, {Blanco},
  {Bramante}, {Cappiello}, {Dutra}, {Ebadi}, {Engel}, {Kolb}, {Harding},
  {Kumar}, {Krnjaic}, {Lang}, {Leane}, {Lehmann}, {Li}, {Long}, {Mohlabeng},
  {Olcina}, {Pueschel}, {Rodd}, {Rott}, {Sengupta}, {Shakya}, {Walsworth}, \&
  {Westerdale}}]{2022arXiv220306508C}
{Carney}, D., {Raj}, N., {Bai}, Y., {et~al.} 2022, arXiv e-prints,
  arXiv:2203.06508.
\newblock \doarXiv{2203.06508}

\bibitem[{Chen {et~al.}(2009)Chen, Nojiri, Takahashi, \&
  Yanagida}]{Chen:2008qs}
Chen, C.-R., Nojiri, M.~M., Takahashi, F., \& Yanagida, T.~T. 2009, Prog.
  Theor. Phys., 122, 553, \dodoi{10.1143/PTP.122.553}

\bibitem[{{de Jong} {et~al.}(2019){de Jong}, {Agertz}, {Berbel}, {Aird},
  {Alexander}, {Amarsi}, {Anders}, {Andrae}, {Ansarinejad}, {Ansorge},
  {Antilogus}, {Anwand-Heerwart}, {Arentsen}, {Arnadottir}, {Asplund}, {Auger},
  {Azais}, {Baade}, {Baker}, {Baker}, {Balbinot}, {Baldry}, {Banerji},
  {Barden}, {Barklem}, {Barth{\'e}l{\'e}my-Mazot}, {Battistini}, {Bauer},
  {Bell}, {Bellido-Tirado}, {Bellstedt}, {Belokurov}, {Bensby}, {Bergemann},
  {Bestenlehner}, {Bielby}, {Bilicki}, {Blake}, {Bland-Hawthorn}, {Boeche},
  {Boland}, {Boller}, {Bongard}, {Bongiorno}, {Bonifacio}, {Boudon}, {Brooks},
  {Brown}, {Brown}, {Br{\"u}ggen}, {Brynnel}, {Brzeski}, {Buchert},
  {Buschkamp}, {Caffau}, {Caillier}, {Carrick}, {Casagrande}, {Case}, {Casey},
  {Cesarini}, {Cescutti}, {Chapuis}, {Chiappini}, {Childress}, {Christlieb},
  {Church}, {Cioni}, {Cluver}, {Colless}, {Collett}, {Comparat}, {Cooper},
  {Couch}, {Courbin}, {Croom}, {Croton}, {Daguis{\'e}}, {Dalton}, {Davies},
  {Davis}, {de Laverny}, {Deason}, {Dionies}, {Disseau}, {Doel}, {D{\"o}scher},
  {Driver}, {Dwelly}, {Eckert}, {Edge}, {Edvardsson}, {Youssoufi}, {Elhaddad},
  {Enke}, {Erfanianfar}, {Farrell}, {Fechner}, {Feiz}, {Feltzing}, {Ferreras},
  {Feuerstein}, {Feuillet}, {Finoguenov}, {Ford}, {Fotopoulou}, {Fouesneau},
  {Frenk}, {Frey}, {Gaessler}, {Geier}, {Gentile Fusillo}, {Gerhard},
  {Giannantonio}, {Giannone}, {Gibson}, {Gillingham},
  {Gonz{\'a}lez-Fern{\'a}ndez}, {Gonzalez-Solares}, {Gottloeber}, {Gould},
  {Grebel}, {Gueguen}, {Guiglion}, {Haehnelt}, {Hahn}, {Hansen}, {Hartman},
  {Hauptner}, {Hawkins}, {Haynes}, {Haynes}, {Heiter}, {Helmi}, {Aguayo},
  {Hewett}, {Hinton}, {Hobbs}, {Hoenig}, {Hofman}, {Hook}, {Hopgood},
  {Hopkins}, {Hourihane}, {Howes}, {Howlett}, {Huet}, {Irwin}, {Iwert},
  {Jablonka}, {Jahn}, {Jahnke}, {Jarno}, {Jin}, {Jofre}, {Johl}, {Jones},
  {J{\"o}nsson}, {Jordan}, {Karovicova}, {Khalatyan}, {Kelz}, {Kennicutt},
  {King}, {Kitaura}, {Klar}, {Klauser}, {Kneib}, {Koch}, {Koposov},
  {Kordopatis}, {Korn}, {Kosmalski}, {Kotak}, {Kovalev}, {Kreckel}, {Kripak},
  {Krumpe}, {Kuijken}, {Kunder}, {Kushniruk}, {Lam}, {Lamer}, {Laurent},
  {Lawrence}, {Lehmitz}, {Lemasle}, {Lewis}, {Li}, {Lidman}, {Lind}, {Liske},
  {Lizon}, {Loveday}, {Ludwig}, {McDermid}, {Maguire}, {Mainieri}, {Mali},
  {Mandel}, {Mandel}, {Mannering}, {Martell}, {Martinez Delgado}, {Matijevic},
  {McGregor}, {McMahon}, {McMillan}, {Mena}, {Merloni}, {Meyer}, {Michel},
  {Micheva}, {Migniau}, {Minchev}, {Monari}, {Muller}, {Murphy},
  {Muthukrishna}, {Nandra}, {Navarro}, {Ness}, {Nichani}, {Nichol}, {Nicklas},
  {Niederhofer}, {Norberg}, {Obreschkow}, {Oliver}, {Owers}, {Pai},
  {Pankratow}, {Parkinson}, {Paschke}, {Paterson}, {Pecontal}, {Parry},
  {Phillips}, {Pillepich}, {Pinard}, {Pirard}, {Piskunov}, {Plank},
  {Pl{\"u}schke}, {Pons}, {Popesso}, {Power}, {Pragt}, {Pramskiy}, {Pryer},
  {Quattri}, {Queiroz}, {Quirrenbach}, {Rahurkar}, {Raichoor}, {Ramstedt},
  {Rau}, {Recio-Blanco}, {Reiss}, {Renaud}, {Revaz}, {Rhode}, {Richard},
  {Richter}, {Rix}, {Robotham}, {Roelfsema}, {Romaniello}, {Rosario},
  {Rothmaier}, {Roukema}, {Ruchti}, {Rupprecht}, {Rybizki}, {Ryde}, {Saar},
  {Sadler}, {Sahl{\'e}n}, {Salvato}, {Sassolas}, {Saunders}, {Saviauk},
  {Sbordone}, {Schmidt}, {Schnurr}, {Scholz}, {Schwope}, {Seifert}, {Shanks},
  {Sheinis}, {Sivov}, {Sk{\'u}lad{\'o}ttir}, {Smartt}, {Smedley}, {Smith},
  {Smith}, {Sorce}, {Spitler}, {Starkenburg}, {Steinmetz}, {Stilz}, {Storm},
  {Sullivan}, {Sutherland}, {Swann}, {Tamone}, {Taylor}, {Teillon}, {Tempel},
  {ter Horst}, {Thi}, {Tolstoy}, {Trager}, {Traven}, {Tremblay}, {Tresse},
  {Valentini}, {van de Weygaert}, {van den Ancker}, {Veljanoski}, {Venkatesan},
  {Wagner}, {Wagner}, {Walcher}, {Waller}, {Walton}, {Wang}, {Winkler},
  {Wisotzki}, {Worley}, {Worseck}, {Xiang}, {Xu}, {Yong}, {Zhao}, {Zheng},
  {Zscheyge}, \& {Zucker}}]{2019Msngr.175....3D}
{de Jong}, R.~S., {Agertz}, O., {Berbel}, A.~A., {et~al.} 2019, The Messenger,
  175, 3, \dodoi{10.18727/0722-6691/5117}

\bibitem[{Dicus {et~al.}(1978)Dicus, Kolb, \& Teplitz}]{Dicus:1977qy}
Dicus, D.~A., Kolb, E.~W., \& Teplitz, V.~L. 1978, Astrophys. J., 221, 327,
  \dodoi{10.1086/156031}

\bibitem[{Ellis {et~al.}(1984)Ellis, Kim, \& Nanopoulos}]{Ellis:1984eq}
Ellis, J.~R., Kim, J.~E., \& Nanopoulos, D.~V. 1984, Phys. Lett. B, 145, 181,
  \dodoi{10.1016/0370-2693(84)90334-4}

\bibitem[{{Ellis} {et~al.}(2017){Ellis}, {Bland-Hawthorn}, {Bremer},
  {Brinchmann}, {Guzzo}, {Richard}, {Rix}, {Tolstoy}, \&
  {Watson}}]{2017arXiv170101976E}
{Ellis}, R.~S., {Bland-Hawthorn}, J., {Bremer}, M., {et~al.} 2017, arXiv
  e-prints, arXiv:1701.01976.
\newblock \doarXiv{1701.01976}

\bibitem[{{Emami} {et~al.}(2021){Emami}, {Genel}, {Hernquist}, {Alcock},
  {Bose}, {Weinberger}, {Vogelsberger}, {Marinacci}, {Loeb}, {Torrey}, \&
  {Forbes}}]{2021ApJ...913...36E}
{Emami}, R., {Genel}, S., {Hernquist}, L., {et~al.} 2021, \apj, 913, 36,
  \dodoi{10.3847/1538-4357/abf147}

\bibitem[{{Erkal} {et~al.}(2021){Erkal}, {Deason}, {Belokurov}, {Xue},
  {Koposov}, {Bird}, {Liu}, {Simion}, {Yang}, {Zhang}, \&
  {Zhao}}]{2021MNRAS.506.2677E}
{Erkal}, D., {Deason}, A.~J., {Belokurov}, V., {et~al.} 2021, \mnras, 506,
  2677, \dodoi{10.1093/mnras/stab1828}

\bibitem[{{Feng}(2010)}]{2010ARA&A..48..495F}
{Feng}, J.~L. 2010, \araa, 48, 495, \dodoi{10.1146/annurev-astro-082708-101659}

\bibitem[{{Fernandez-Martinez} {et~al.}(2021){Fernandez-Martinez}, {Pierre},
  {Pinsard}, \& {Rosauro-Alcaraz}}]{2021EPJC...81..954F}
{Fernandez-Martinez}, E., {Pierre}, M., {Pinsard}, E., \& {Rosauro-Alcaraz}, S.
  2021, European Physical Journal C, 81, 954,
  \dodoi{10.1140/epjc/s10052-021-09760-y}

\bibitem[{{Gaia Collaboration} {et~al.}(2018){Gaia Collaboration}, {Katz},
  {Antoja}, {Romero-G{\'o}mez}, {Drimmel}, {Reyl{\'e}}, {Seabroke}, {Soubiran},
  {Babusiaux}, {Di Matteo}, {Figueras}, {Poggio}, {Robin}, {Evans}, {Brown},
  {Vallenari}, {Prusti}, {de Bruijne}, {Bailer-Jones}, {Biermann}, {Eyer},
  {Jansen}, {Jordi}, {Klioner}, {Lammers}, {Lindegren}, {Luri}, {Mignard},
  {Panem}, {Pourbaix}, {Randich}, {Sartoretti}, {Siddiqui}, {van Leeuwen},
  {Walton}, {Arenou}, {Bastian}, {Cropper}, {Lattanzi}, {Bakker}, {Cacciari},
  {Casta n}, {Chaoul}, {Cheek}, {De Angeli}, {Fabricius}, {Guerra}, {Holl},
  {Masana}, {Messineo}, {Mowlavi}, {Nienartowicz}, {Panuzzo}, {Portell},
  {Riello}, {Tanga}, {Th{\'e}venin}, {Gracia-Abril}, {Comoretto},
  {Garcia-Reinaldos}, {Teyssier}, {Altmann}, {Andrae}, {Audard},
  {Bellas-Velidis}, {Benson}, {Berthier}, {Blomme}, {Burgess}, {Busso},
  {Carry}, {Cellino}, {Clementini}, {Clotet}, {Creevey}, {Davidson}, {De
  Ridder}, {Delchambre}, {Dell'Oro}, {Ducourant},
  {Fern{\'a}ndez-Hern{\'a}ndez}, {Fouesneau}, {Fr{\'e}mat}, {Galluccio},
  {Garc{\'\i}a-Torres}, {Gonz{\'a}lez-N{\'u}{\~n}ez}, {Gonz{\'a}lez-Vidal},
  {Gosset}, {Guy}, {Halbwachs}, {Hambly}, {Harrison}, {Hern{\'a}ndez},
  {Hestroffer}, {Hodgkin}, {Hutton}, {Jasniewicz}, {Jean-Antoine-Piccolo},
  {Jordan}, {Korn}, {Krone-Martins}, {Lanzafame}, {Lebzelter}, {L{\"o}ffler},
  {Manteiga}, {Marrese}, {Mart{\'\i}n-Fleitas}, {Moitinho}, {Mora}, {Muinonen},
  {Osinde}, {Pancino}, {Pauwels}, {Petit}, {Recio-Blanco}, {Richards},
  {Rimoldini}, {Sarro}, {Siopis}, {Smith}, {Sozzetti}, {S{\"u}veges}, {Torra},
  {van Reeven}, {Abbas}, {Abreu Aramburu}, {Accart}, {Aerts}, {Altavilla},
  {{\'A}lvarez}, {Alvarez}, {Alves}, {Anderson}, {Andrei}, {Anglada Varela},
  {Antiche}, {Arcay}, {Astraatmadja}, {Bach}, {Baker},
  {Balaguer-N{\'u}{\~n}ez}, {Balm}, {Barache}, {Barata}, {Barbato}, {Barblan},
  {Barklem}, {Barrado}, {Barros}, {Barstow}, {Bartholom{\'e} Mu{\~n}oz},
  {Bassilana}, {Becciani}, {Bellazzini}, {Berihuete}, {Bertone}, {Bianchi},
  {Bienaym{\'e}}, {Blanco-Cuaresma}, {Boch}, {Boeche}, {Bombrun}, {Borrachero},
  {Bossini}, {Bouquillon}, {Bourda}, {Bragaglia}, {Bramante}, {Breddels},
  {Bressan}, {Brouillet}, {Br{\"u}semeister}, {Brugaletta}, {Bucciarelli},
  {Burlacu}, {Busonero}, {Butkevich}, {Buzzi}, {Caffau}, {Cancelliere},
  {Cannizzaro}, {Cantat-Gaudin}, {Carballo}, {Carlucci}, {Carrasco},
  {Casamiquela}, {Castellani}, {Castro-Ginard}, {Charlot}, {Chemin},
  {Chiavassa}, {Cocozza}, {Costigan}, {Cowell}, {Crifo}, {Crosta}, {Crowley},
  {Cuypers}, {Dafonte}, {Damerdji}, {Dapergolas}, {David}, {David}, {de
  Laverny}, {De Luise}, {De March}, {de Souza}, {de Torres}, {Debosscher}, {del
  Pozo}, {Delbo}, {Delgado}, {Delgado}, {Diakite}, {Diener}, {Distefano},
  {Dolding}, {Drazinos}, {Dur{\'a}n}, {Edvardsson}, {Enke}, {Eriksson},
  {Esquej}, {Eynard Bontemps}, {Fabre}, {Fabrizio}, {Faigler}, {Falc a},
  {Farr{\`a}s Casas}, {Federici}, {Fedorets}, {Fernique}, {Filippi},
  {Findeisen}, {Fonti}, {Fraile}, {Fraser}, {Fr{\'e}zouls}, {Gai}, {Galleti},
  {Garabato}, {Garc{\'\i}a-Sedano}, {Garofalo}, {Garralda}, {Gavel}, {Gavras},
  {Gerssen}, {Geyer}, {Giacobbe}, {Gilmore}, {Girona}, {Giuffrida}, {Glass},
  {Gomes}, {Granvik}, {Gueguen}, {Guerrier}, {Guiraud}, {Guti{\'e}}, {Haigron},
  {Hatzidimitriou}, {Hauser}, {Haywood}, {Heiter}, {Helmi}, {Heu}, {Hilger},
  {Hobbs}, {Hofmann}, {Holland}, {Huckle}, {Hypki}, {Icardi}, {Jan{\ss}en},
  {Jevardat de Fombelle}, {Jonker}, {Juh{\'a}sz}, {Julbe}, {Karampelas},
  {Kewley}, {Klar}, {Kochoska}, {Kohley}, {Kolenberg}, {Kontizas}, {Kontizas},
  {Koposov}, {Kordopatis}, {Kostrzewa-Rutkowska}, {Koubsky}, {Lambert},
  {Lanza}, {Lasne}, {Lavigne}, {Le Fustec}, {Le Poncin-Lafitte}, {Lebreton},
  {Leccia}, {Leclerc}, {Lecoeur-Taibi}, {Lenhardt}, {Leroux}, {Liao}, {Licata},
  {Lindstr{\o}m}, {Lister}, {Livanou}, {Lobel}, {L{\'o}pez}, {Managau}, {Mann},
  {Mantelet}, {Marchal}, {Marchant}, {Marconi}, {Marinoni}, {Marschalk{\'o}},
  {Marshall}, {Martino}, {Marton}, {Mary}, {Massari}, {Matijevi{\v{c}}},
  {Mazeh}, {McMillan}, {Messina}, {Michalik}, {Millar}, {Molina}, {Molinaro},
  {Moln{\'a}r}, {Montegriffo}, {Mor}, {Morbidelli}, {Morel}, {Morris},
  {Mulone}, {Muraveva}, {Musella}, {Nelemans}, {Nicastro}, {Noval},
  {O'Mullane}, {Ord{\'e}novic}, {Ord{\'o}{\~n}ez-Blanco}, {Osborne}, {Pagani},
  {Pagano}, {Pailler}, {Palacin}, {Palaversa}, {Panahi}, {Pawlak},
  {Piersimoni}, {Pineau}, {Plachy}, {Plum}, {Poujoulet}, {Pr{\v{s}}a},
  {Pulone}, {Racero}, {Ragaini}, {Rambaux}, {Ramos-Lerate}, {Regibo}, {Riclet},
  {Ripepi}, {Riva}, {Rivard}, {Rixon}, {Roegiers}, {Roelens}, {Rowell},
  {Royer}, {Ruiz-Dern}, {Sadowski}, {Sagrist{\`a} Sell{\'e}s}, {Sahlmann},
  {Salgado}, {Salguero}, {Sanna}, {Santana-Ros}, {Sarasso}, {Savietto},
  {Schultheis}, {Sciacca}, {Segol}, {Segovia}, {S{\'e}gransan}, {Shih},
  {Siltala}, {Silva}, {Smart}, {Smith}, {Solano}, {Solitro}, {Sordo}, {Soria
  Nieto}, {Souchay}, {Spagna}, {Spoto}, {Stampa}, {Steele},
  {Steidelm{\"u}ller}, {Stephenson}, {Stoev}, {Suess}, {Surdej}, {Szabados},
  {Szegedi-Elek}, {Tapiador}, {Taris}, {Tauran}, {Taylor}, {Teixeira},
  {Terrett}, {Teyssandier}, {Thuillot}, {Titarenko}, {Torra Clotet}, {Turon},
  {Ulla}, {Utrilla}, {Uzzi}, {Vaillant}, {Valentini}, {Valette}, {van Elteren},
  {Van Hemelryck}, {van Leeuwen}, {Vaschetto}, {Vecchiato}, {Veljanoski},
  {Viala}, {Vicente}, {Vogt}, {von Essen}, {Voss}, {Votruba}, {Voutsinas},
  {Walmsley}, {Weiler}, {Wertz}, {Wevers}, {Wyrzykowski}, {Yoldas},
  {{\v{Z}}erjal}, {Ziaeepour}, {Zorec}, {Zschocke}, {Zucker}, {Zurbach}, \&
  {Zwitter}}]{2018A&A...616A..11G}
{Gaia Collaboration}, {Katz}, D., {Antoja}, T., {et~al.} 2018, \aap, 616, A11,
  \dodoi{10.1051/0004-6361/201832865}

\bibitem[{{Gaia Collaboration} {et~al.}(2021){Gaia Collaboration}, {Brown},
  {Vallenari}, {Prusti}, {de Bruijne}, {Babusiaux}, {Biermann}, {Creevey},
  {Evans}, {Eyer}, {Hutton}, {Jansen}, {Jordi}, {Klioner}, {Lammers},
  {Lindegren}, {Luri}, {Mignard}, {Panem}, {Pourbaix}, {Randich}, {Sartoretti},
  {Soubiran}, {Walton}, {Arenou}, {Bailer-Jones}, {Bastian}, {Cropper},
  {Drimmel}, {Katz}, {Lattanzi}, {van Leeuwen}, {Bakker}, {Cacciari},
  {Casta{\~n}eda}, {De Angeli}, {Ducourant}, {Fabricius}, {Fouesneau},
  {Fr{\'e}mat}, {Guerra}, {Guerrier}, {Guiraud}, {Jean-Antoine Piccolo},
  {Masana}, {Messineo}, {Mowlavi}, {Nicolas}, {Nienartowicz}, {Pailler},
  {Panuzzo}, {Riclet}, {Roux}, {Seabroke}, {Sordo}, {Tanga}, {Th{\'e}venin},
  {Gracia-Abril}, {Portell}, {Teyssier}, {Altmann}, {Andrae}, {Bellas-Velidis},
  {Benson}, {Berthier}, {Blomme}, {Brugaletta}, {Burgess}, {Busso}, {Carry},
  {Cellino}, {Cheek}, {Clementini}, {Damerdji}, {Davidson}, {Delchambre},
  {Dell'Oro}, {Fern{\'a}ndez-Hern{\'a}ndez}, {Galluccio}, {Garc{\'\i}a-Lario},
  {Garcia-Reinaldos}, {Gonz{\'a}lez-N{\'u}{\~n}ez}, {Gosset}, {Haigron},
  {Halbwachs}, {Hambly}, {Harrison}, {Hatzidimitriou}, {Heiter},
  {Hern{\'a}ndez}, {Hestroffer}, {Hodgkin}, {Holl}, {Jan{\ss}en}, {Jevardat de
  Fombelle}, {Jordan}, {Krone-Martins}, {Lanzafame}, {L{\"o}ffler}, {Lorca},
  {Manteiga}, {Marchal}, {Marrese}, {Moitinho}, {Mora}, {Muinonen}, {Osborne},
  {Pancino}, {Pauwels}, {Petit}, {Recio-Blanco}, {Richards}, {Riello},
  {Rimoldini}, {Robin}, {Roegiers}, {Rybizki}, {Sarro}, {Siopis}, {Smith},
  {Sozzetti}, {Ulla}, {Utrilla}, {van Leeuwen}, {van Reeven}, {Abbas}, {Abreu
  Aramburu}, {Accart}, {Aerts}, {Aguado}, {Ajaj}, {Altavilla}, {{\'A}lvarez},
  {{\'A}lvarez Cid-Fuentes}, {Alves}, {Anderson}, {Anglada Varela}, {Antoja},
  {Audard}, {Baines}, {Baker}, {Balaguer-N{\'u}{\~n}ez}, {Balbinot}, {Balog},
  {Barache}, {Barbato}, {Barros}, {Barstow}, {Bartolom{\'e}}, {Bassilana},
  {Bauchet}, {Baudesson-Stella}, {Becciani}, {Bellazzini}, {Bernet}, {Bertone},
  {Bianchi}, {Blanco-Cuaresma}, {Boch}, {Bombrun}, {Bossini}, {Bouquillon},
  {Bragaglia}, {Bramante}, {Breedt}, {Bressan}, {Brouillet}, {Bucciarelli},
  {Burlacu}, {Busonero}, {Butkevich}, {Buzzi}, {Caffau}, {Cancelliere},
  {C{\'a}novas}, {Cantat-Gaudin}, {Carballo}, {Carlucci}, {Carnerero},
  {Carrasco}, {Casamiquela}, {Castellani}, {Castro-Ginard}, {Castro Sampol},
  {Chaoul}, {Charlot}, {Chemin}, {Chiavassa}, {Cioni}, {Comoretto}, {Cooper},
  {Cornez}, {Cowell}, {Crifo}, {Crosta}, {Crowley}, {Dafonte}, {Dapergolas},
  {David}, {David}, {de Laverny}, {De Luise}, {De March}, {De Ridder}, {de
  Souza}, {de Teodoro}, {de Torres}, {del Peloso}, {del Pozo}, {Delbo},
  {Delgado}, {Delgado}, {Delisle}, {Di Matteo}, {Diakite}, {Diener},
  {Distefano}, {Dolding}, {Eappachen}, {Edvardsson}, {Enke}, {Esquej}, {Fabre},
  {Fabrizio}, {Faigler}, {Fedorets}, {Fernique}, {Fienga}, {Figueras},
  {Fouron}, {Fragkoudi}, {Fraile}, {Franke}, {Gai}, {Garabato},
  {Garcia-Gutierrez}, {Garc{\'\i}a-Torres}, {Garofalo}, {Gavras}, {Gerlach},
  {Geyer}, {Giacobbe}, {Gilmore}, {Girona}, {Giuffrida}, {Gomel}, {Gomez},
  {Gonzalez-Santamaria}, {Gonz{\'a}lez-Vidal}, {Granvik},
  {Guti{\'e}rrez-S{\'a}nchez}, {Guy}, {Hauser}, {Haywood}, {Helmi}, {Hidalgo},
  {Hilger}, {H{\l}adczuk}, {Hobbs}, {Holland}, {Huckle}, {Jasniewicz},
  {Jonker}, {Juaristi Campillo}, {Julbe}, {Karbevska}, {Kervella}, {Khanna},
  {Kochoska}, {Kontizas}, {Kordopatis}, {Korn}, {Kostrzewa-Rutkowska},
  {Kruszy{\'n}ska}, {Lambert}, {Lanza}, {Lasne}, {Le Campion}, {Le Fustec},
  {Lebreton}, {Lebzelter}, {Leccia}, {Leclerc}, {Lecoeur-Taibi}, {Liao},
  {Licata}, {Lindstr{\o}m}, {Lister}, {Livanou}, {Lobel}, {Madrero Pardo},
  {Managau}, {Mann}, {Marchant}, {Marconi}, {Marcos Santos}, {Marinoni},
  {Marocco}, {Marshall}, {Martin Polo}, {Mart{\'\i}n-Fleitas}, {Masip},
  {Massari}, {Mastrobuono-Battisti}, {Mazeh}, {McMillan}, {Messina},
  {Michalik}, {Millar}, {Mints}, {Molina}, {Molinaro}, {Moln{\'a}r},
  {Montegriffo}, {Mor}, {Morbidelli}, {Morel}, {Morris}, {Mulone}, {Munoz},
  {Muraveva}, {Murphy}, {Musella}, {Noval}, {Ord{\'e}novic}, {Orr{\`u}},
  {Osinde}, {Pagani}, {Pagano}, {Palaversa}, {Palicio}, {Panahi}, {Pawlak},
  {Pe{\~n}alosa Esteller}, {Penttil{\"a}}, {Piersimoni}, {Pineau}, {Plachy},
  {Plum}, {Poggio}, {Poretti}, {Poujoulet}, {Pr{\v{s}}a}, {Pulone}, {Racero},
  {Ragaini}, {Rainer}, {Raiteri}, {Rambaux}, {Ramos}, {Ramos-Lerate}, {Re
  Fiorentin}, {Regibo}, {Reyl{\'e}}, {Ripepi}, {Riva}, {Rixon}, {Robichon},
  {Robin}, {Roelens}, {Rohrbasser}, {Romero-G{\'o}mez}, {Rowell}, {Royer},
  {Rybicki}, {Sadowski}, {Sagrist{\`a} Sell{\'e}s}, {Sahlmann}, {Salgado},
  {Salguero}, {Samaras}, {Sanchez Gimenez}, {Sanna}, {Santove{\~n}a},
  {Sarasso}, {Schultheis}, {Sciacca}, {Segol}, {Segovia}, {S{\'e}gransan},
  {Semeux}, {Shahaf}, {Siddiqui}, {Siebert}, {Siltala}, {Slezak}, {Smart},
  {Solano}, {Solitro}, {Souami}, {Souchay}, {Spagna}, {Spoto}, {Steele},
  {Steidelm{\"u}ller}, {Stephenson}, {S{\"u}veges}, {Szabados}, {Szegedi-Elek},
  {Taris}, {Tauran}, {Taylor}, {Teixeira}, {Thuillot}, {Tonello}, {Torra},
  {Torra}, {Turon}, {Unger}, {Vaillant}, {van Dillen}, {Vanel}, {Vecchiato},
  {Viala}, {Vicente}, {Voutsinas}, {Weiler}, {Wevers}, {Wyrzykowski}, {Yoldas},
  {Yvard}, {Zhao}, {Zorec}, {Zucker}, {Zurbach}, \&
  {Zwitter}}]{2021A&A...649A...1G}
{Gaia Collaboration}, {Brown}, A.~G.~A., {Vallenari}, A., {et~al.} 2021, \aap,
  649, A1, \dodoi{10.1051/0004-6361/202039657}

\bibitem[{{Helmi}(2008)}]{2008A&ARv..15..145H}
{Helmi}, A. 2008, \aapr, 15, 145, \dodoi{10.1007/s00159-008-0009-6}

\bibitem[{{Helmi} {et~al.}(2019){Helmi}, {Irwin}, {Deason}, {Balbinot},
  {Belokurov}, {Bland-Hawthorn}, {Christlieb}, {Cioni}, {Feltzing}, {Grebel},
  {Kordopatis}, {Starkenburg}, {Walton}, \& {Worley}}]{2019Msngr.175...23H}
{Helmi}, A., {Irwin}, M., {Deason}, A., {et~al.} 2019, The Messenger, 175, 23,
  \dodoi{10.18727/0722-6691/5120}

\bibitem[{{Hopkins} {et~al.}(2018){Hopkins}, {Wetzel}, {Kere{\v{s}}},
  {Faucher-Gigu{\`e}re}, {Quataert}, {Boylan-Kolchin}, {Murray}, {Hayward},
  {Garrison-Kimmel}, {Hummels}, {Feldmann}, {Torrey}, {Ma},
  {Angl{\'e}s-Alc{\'a}zar}, {Su}, {Orr}, {Schmitz}, {Escala}, {Sanderson},
  {Grudi{\'c}}, {Hafen}, {Kim}, {Fitts}, {Bullock}, {Wheeler}, {Chan},
  {Elbert}, \& {Narayanan}}]{2018MNRAS.480..800H}
{Hopkins}, P.~F., {Wetzel}, A., {Kere{\v{s}}}, D., {et~al.} 2018, \mnras, 480,
  800, \dodoi{10.1093/mnras/sty1690}

\bibitem[{{Hubert} {et~al.}(2021){Hubert}, {Schneider}, {Potter}, {Stadel}, \&
  {Giri}}]{2021JCAP...10..040H}
{Hubert}, J., {Schneider}, A., {Potter}, D., {Stadel}, J., \& {Giri}, S.~K.
  2021, \jcap, 2021, 040, \dodoi{10.1088/1475-7516/2021/10/040}

\bibitem[{Ibarra \& Tran(2009)}]{Ibarra:2008jk}
Ibarra, A., \& Tran, D. 2009, JCAP, 02, 021,
  \dodoi{10.1088/1475-7516/2009/02/021}

\bibitem[{Jeltema \& Profumo(2015)}]{Jeltema:2014qfa}
Jeltema, T.~E., \& Profumo, S. 2015, Mon. Not. Roy. Astron. Soc., 450, 2143,
  \dodoi{10.1093/mnras/stv768}

\bibitem[{{Johnston} {et~al.}(2008){Johnston}, {Bullock}, {Sharma}, {Font},
  {Robertson}, \& {Leitner}}]{2008ApJ...689..936J}
{Johnston}, K.~V., {Bullock}, J.~S., {Sharma}, S., {et~al.} 2008, \apj, 689,
  936, \dodoi{10.1086/592228}

\bibitem[{{Kahlhoefer}(2017)}]{2017IJMPA..3230006K}
{Kahlhoefer}, F. 2017, International Journal of Modern Physics A, 32, 1730006,
  \dodoi{10.1142/S0217751X1730006X}

\bibitem[{Kawasaki {et~al.}(2008)Kawasaki, Nakayama, \&
  Senami}]{Kawasaki:2007mk}
Kawasaki, M., Nakayama, K., \& Senami, M. 2008, JCAP, 03, 009,
  \dodoi{10.1088/1475-7516/2008/03/009}

\bibitem[{{Khanna} {et~al.}(2019{\natexlab{a}}){Khanna}, {Sharma},
  {Bland-Hawthorn}, {Hayden}, {Nataf}, {Ting}, {Kos}, {Martell}, {Zwitter}, {De
  Silva}, {Asplund}, {Buder}, {Duong}, {Lin}, {Simpson}, {Anguiano}, {Horner},
  {Kafle}, {Lewis}, {Nordlander}, {Wyse}, {Wittenmyer}, \&
  {Zucker}}]{2019MNRAS.482.4215K}
{Khanna}, S., {Sharma}, S., {Bland-Hawthorn}, J., {et~al.} 2019{\natexlab{a}},
  \mnras, 482, 4215, \dodoi{10.1093/mnras/sty2924}

\bibitem[{{Khanna} {et~al.}(2019{\natexlab{b}}){Khanna}, {Sharma},
  {Tepper-Garcia}, {Bland-Hawthorn}, {Hayden}, {Asplund}, {Buder}, {Chen}, {De
  Silva}, {Freeman}, {Kos}, {Lewis}, {Lin}, {Martell}, {Simpson}, {Nordlander},
  {Stello}, {Ting}, {Zucker}, \& {Zwitter}}]{2019MNRAS.489.4962K}
{Khanna}, S., {Sharma}, S., {Tepper-Garcia}, T., {et~al.} 2019{\natexlab{b}},
  \mnras, 489, 4962, \dodoi{10.1093/mnras/stz2462}

\bibitem[{Kim {et~al.}(2022)Kim, Lopez-Fogliani, Perez, \&
  de~Austri}]{Kim:2022gpl}
Kim, J.~S., Lopez-Fogliani, D.~E., Perez, A.~D., \& de~Austri, R.~R. 2022,
  arXiv e-prints.
\newblock \doarXiv{2206.04715}

\bibitem[{{Laporte} {et~al.}(2018){Laporte}, {Johnston}, {G{\'o}mez},
  {Garavito-Camargo}, \& {Besla}}]{2018MNRAS.481..286L}
{Laporte}, C. F.~P., {Johnston}, K.~V., {G{\'o}mez}, F.~A., {Garavito-Camargo},
  N., \& {Besla}, G. 2018, \mnras, 481, 286, \dodoi{10.1093/mnras/sty1574}

\bibitem[{{Laporte} {et~al.}(2019){Laporte}, {Minchev}, {Johnston}, \&
  {G{\'o}mez}}]{2019MNRAS.485.3134L}
{Laporte}, C. F.~P., {Minchev}, I., {Johnston}, K.~V., \& {G{\'o}mez}, F.~A.
  2019, \mnras, 485, 3134, \dodoi{10.1093/mnras/stz583}

\bibitem[{{Liu} {et~al.}(2017){Liu}, {Chen}, \& {Ji}}]{2017NatPh..13..212L}
{Liu}, J., {Chen}, X., \& {Ji}, X. 2017, Nature Physics, 13, 212,
  \dodoi{10.1038/nphys4039}

\bibitem[{{Loeb}(2022)}]{2022RNAAS...6...26L}
{Loeb}, A. 2022, Research Notes of the American Astronomical Society, 6, 26,
  \dodoi{10.3847/2515-5172/ac5185}

\bibitem[{{Majewski} {et~al.}(2017){Majewski}, {Schiavon}, {Frinchaboy},
  {Allende Prieto}, {Barkhouser}, {Bizyaev}, {Blank}, {Brunner}, {Burton},
  {Carrera}, {Chojnowski}, {Cunha}, {Epstein}, {Fitzgerald}, {Garc{\'\i}a
  P{\'e}rez}, {Hearty}, {Henderson}, {Holtzman}, {Johnson}, {Lam}, {Lawler},
  {Maseman}, {M{\'e}sz{\'a}ros}, {Nelson}, {Nguyen}, {Nidever}, {Pinsonneault},
  {Shetrone}, {Smee}, {Smith}, {Stolberg}, {Skrutskie}, {Walker}, {Wilson},
  {Zasowski}, {Anders}, {Basu}, {Beland}, {Blanton}, {Bovy}, {Brownstein},
  {Carlberg}, {Chaplin}, {Chiappini}, {Eisenstein}, {Elsworth}, {Feuillet},
  {Fleming}, {Galbraith-Frew}, {Garc{\'\i}a}, {Garc{\'\i}a-Hern{\'a}ndez},
  {Gillespie}, {Girardi}, {Gunn}, {Hasselquist}, {Hayden}, {Hekker}, {Ivans},
  {Kinemuchi}, {Klaene}, {Mahadevan}, {Mathur}, {Mosser}, {Muna}, {Munn},
  {Nichol}, {O'Connell}, {Parejko}, {Robin}, {Rocha-Pinto}, {Schultheis},
  {Serenelli}, {Shane}, {Silva Aguirre}, {Sobeck}, {Thompson}, {Troup},
  {Weinberg}, \& {Zamora}}]{2017AJ....154...94M}
{Majewski}, S.~R., {Schiavon}, R.~P., {Frinchaboy}, P.~M., {et~al.} 2017, \aj,
  154, 94, \dodoi{10.3847/1538-3881/aa784d}

\bibitem[{{Mau} {et~al.}(2022){Mau}, {Nadler}, {Wechsler}, {Drlica-Wagner},
  {Bechtol}, {Green}, {Huterer}, {Li}, {Mao}, {Mart{\'\i}nez-V{\'a}zquez},
  {McNanna}, {Mutlu-Pakdil}, {Pace}, {Peter}, {Riley}, {Strigari}, {Wang},
  {Aguena}, {Allam}, {Annis}, {Bacon}, {Bertin}, {Bocquet}, {Brooks}, {Burke},
  {Carnero Rosell}, {Carrasco Kind}, {Carretero}, {Costanzi}, {Crocce},
  {Pereira}, {Davis}, {De Vicente}, {Desai}, {Doel}, {Ferrero}, {Flaugher},
  {Frieman}, {Garc{\'\i}a-Bellido}, {Gatti}, {Giannini}, {Gruen}, {Gruendl},
  {Gschwend}, {Gutierrez}, {Hinton}, {Hollowood}, {Honscheid}, {James},
  {Kuehn}, {Lahav}, {Maia}, {Marshall}, {Miquel}, {Mohr}, {Morgan}, {Ogando},
  {Paz-Chinch{\'o}n}, {Pieres}, {Rodriguez-Monroy}, {Sanchez}, {Scarpine},
  {Serrano}, {Sevilla-Noarbe}, {Suchyta}, {Tarle}, {To}, {Tucker}, \&
  {Weller}}]{2022arXiv220111740M}
{Mau}, S., {Nadler}, E.~O., {Wechsler}, R.~H., {et~al.} 2022, arXiv e-prints,
  arXiv:2201.11740.
\newblock \doarXiv{2201.11740}

\bibitem[{{Pandey} {et~al.}(2020){Pandey}, {Karwal}, \&
  {Das}}]{2020JCAP...07..026P}
{Pandey}, K.~L., {Karwal}, T., \& {Das}, S. 2020, \jcap, 2020, 026,
  \dodoi{10.1088/1475-7516/2020/07/026}

\bibitem[{{Peter} \& {Benson}(2010)}]{2010PhRvD..82l3521P}
{Peter}, A. H.~G., \& {Benson}, A.~J. 2010, \prd, 82, 123521,
  \dodoi{10.1103/PhysRevD.82.123521}

\bibitem[{{Pontzen} \& {Governato}(2012)}]{2012MNRAS.421.3464P}
{Pontzen}, A., \& {Governato}, F. 2012, \mnras, 421, 3464,
  \dodoi{10.1111/j.1365-2966.2012.20571.x}

\bibitem[{Poulin {et~al.}(2016)Poulin, Serpico, \&
  Lesgourgues}]{Poulin:2016nat}
Poulin, V., Serpico, P.~D., \& Lesgourgues, J. 2016, JCAP, 08, 036,
  \dodoi{10.1088/1475-7516/2016/08/036}

\bibitem[{{Poulin} {et~al.}(2021){Poulin}, {Smith}, \&
  {Bartlett}}]{2021PhRvD.104l3550P}
{Poulin}, V., {Smith}, T.~L., \& {Bartlett}, A. 2021, \prd, 104, 123550,
  \dodoi{10.1103/PhysRevD.104.123550}

\bibitem[{{Poulin} {et~al.}(2019){Poulin}, {Smith}, {Karwal}, \&
  {Kamionkowski}}]{2019PhRvL.122v1301P}
{Poulin}, V., {Smith}, T.~L., {Karwal}, T., \& {Kamionkowski}, M. 2019, \prl,
  122, 221301, \dodoi{10.1103/PhysRevLett.122.221301}

\bibitem[{Riemer-S\o{}rensen(2016)}]{Riemer-Sorensen:2014yda}
Riemer-S\o{}rensen, S. 2016, Astron. Astrophys., 590, A71,
  \dodoi{10.1051/0004-6361/201527278}

\bibitem[{{Robin} {et~al.}(2003){Robin}, {Reyl{\'e}}, {Derri{\`e}re}, \&
  {Picaud}}]{2003A&A...409..523R}
{Robin}, A.~C., {Reyl{\'e}}, C., {Derri{\`e}re}, S., \& {Picaud}, S. 2003,
  \aap, 409, 523, \dodoi{10.1051/0004-6361:20031117}

\bibitem[{{Sanderson} {et~al.}(2020){Sanderson}, {Wetzel}, {Loebman}, {Sharma},
  {Hopkins}, {Garrison-Kimmel}, {Faucher-Gigu{\`e}re}, {Kere{\v{s}}}, \&
  {Quataert}}]{2020ApJS..246....6S}
{Sanderson}, R.~E., {Wetzel}, A., {Loebman}, S., {et~al.} 2020, \apjs, 246, 6,
  \dodoi{10.3847/1538-4365/ab5b9d}

\bibitem[{{Sharma} {et~al.}(2011{\natexlab{a}}){Sharma}, {Bland-Hawthorn},
  {Johnston}, \& {Binney}}]{2011ApJ...730....3S}
{Sharma}, S., {Bland-Hawthorn}, J., {Johnston}, K.~V., \& {Binney}, J.
  2011{\natexlab{a}}, \apj, 730, 3, \dodoi{10.1088/0004-637X/730/1/3}

\bibitem[{{Sharma} \& {Johnston}(2009)}]{2009ApJ...703.1061S}
{Sharma}, S., \& {Johnston}, K.~V. 2009, \apj, 703, 1061,
  \dodoi{10.1088/0004-637X/703/1/1061}

\bibitem[{{Sharma} {et~al.}(2011{\natexlab{b}}){Sharma}, {Johnston},
  {Majewski}, {Bullock}, \& {Mu{\~n}oz}}]{2011ApJ...728..106S}
{Sharma}, S., {Johnston}, K.~V., {Majewski}, S.~R., {Bullock}, J., \&
  {Mu{\~n}oz}, R.~R. 2011{\natexlab{b}}, \apj, 728, 106,
  \dodoi{10.1088/0004-637X/728/2/106}

\bibitem[{{Sharma} {et~al.}(2010){Sharma}, {Johnston}, {Majewski}, {Mu{\~n}oz},
  {Carlberg}, \& {Bullock}}]{2010ApJ...722..750S}
{Sharma}, S., {Johnston}, K.~V., {Majewski}, S.~R., {et~al.} 2010, \apj, 722,
  750, \dodoi{10.1088/0004-637X/722/1/750}

\bibitem[{{Sharma} \& {Steinmetz}(2006)}]{2006MNRAS.373.1293S}
{Sharma}, S., \& {Steinmetz}, M. 2006, \mnras, 373, 1293,
  \dodoi{10.1111/j.1365-2966.2006.11043.x}

\bibitem[{Silk \& Srednicki(1984)}]{Silk:1984zy}
Silk, J., \& Srednicki, M. 1984, Phys. Rev. Lett., 53, 624,
  \dodoi{10.1103/PhysRevLett.53.624}

\bibitem[{{Vattis} {et~al.}(2019){Vattis}, {Koushiappas}, \&
  {Loeb}}]{2019PhRvD..99l1302V}
{Vattis}, K., {Koushiappas}, S.~M., \& {Loeb}, A. 2019, \prd, 99, 121302,
  \dodoi{10.1103/PhysRevD.99.121302}

\bibitem[{{von Doetinchem} {et~al.}(2020){von Doetinchem}, {Perez}, {Aramaki},
  {Baker}, {Barwick}, {Bird}, {Boezio}, {Boggs}, {Cui}, {Datta}, {Donato},
  {Evoli}, {Fabris}, {Fabbietti}, {Ferronato Bueno}, {Fornengo}, {Fuke},
  {Gerrity}, {Gomez Coral}, {Hailey}, {Hooper}, {Kachelriess}, {Korsmeier},
  {Kozai}, {Lea}, {Li}, {Lowell}, {Manghisoni}, {Moskalenko}, {Munini},
  {Naskret}, {Nelson}, {Ng}, {Nozzoli}, {Oliva}, {Ong}, {Osteria}, {Pierog},
  {Poulin}, {Profumo}, {P{\"o}schl}, {Quinn}, {Re}, {Rogers}, {Ryan},
  {Saffold}, {Sakai}, {Salati}, {Schael}, {Serksnyte}, {Shukla}, {Stoessl},
  {Tjemsland}, {Vannuccini}, {Vecchi}, {Winkler}, {Wright}, {Xiao}, {Xu},
  {Yoshida}, {Zampa}, \& {Zuccon}}]{2020JCAP...08..035V}
{von Doetinchem}, P., {Perez}, K., {Aramaki}, T., {et~al.} 2020, \jcap, 2020,
  035, \dodoi{10.1088/1475-7516/2020/08/035}

\bibitem[{{Wang} {et~al.}(2014){Wang}, {Peter}, {Strigari}, {Zentner}, {Arant},
  {Garrison-Kimmel}, \& {Rocha}}]{2014MNRAS.445..614W}
{Wang}, M.-Y., {Peter}, A. H.~G., {Strigari}, L.~E., {et~al.} 2014, \mnras,
  445, 614, \dodoi{10.1093/mnras/stu1747}

\bibitem[{{Wetzel} {et~al.}(2016){Wetzel}, {Hopkins}, {Kim},
  {Faucher-Gigu{\`e}re}, {Kere{\v{s}}}, \& {Quataert}}]{2016ApJ...827L..23W}
{Wetzel}, A.~R., {Hopkins}, P.~F., {Kim}, J.-h., {et~al.} 2016, \apjl, 827,
  L23, \dodoi{10.3847/2041-8205/827/2/L23}

\bibitem[{{Widrow} {et~al.}(2014){Widrow}, {Barber}, {Chequers}, \&
  {Cheng}}]{2014MNRAS.440.1971W}
{Widrow}, L.~M., {Barber}, J., {Chequers}, M.~H., \& {Cheng}, E. 2014, \mnras,
  440, 1971, \dodoi{10.1093/mnras/stu396}

\bibitem[{Yin {et~al.}(2009)Yin, Yuan, Liu, Zhang, Bi, \& Zhu}]{Yin:2008bs}
Yin, P.-f., Yuan, Q., Liu, J., {et~al.} 2009, Phys. Rev. D, 79, 023512,
  \dodoi{10.1103/PhysRevD.79.023512}

\bibitem[{{Zhao} {et~al.}(2012){Zhao}, {Zhao}, {Chu}, {Jing}, \&
  {Deng}}]{2012RAA....12..723Z}
{Zhao}, G., {Zhao}, Y.-H., {Chu}, Y.-Q., {Jing}, Y.-P., \& {Deng}, L.-C. 2012,
  Research in Astronomy and Astrophysics, 12, 723,
  \dodoi{10.1088/1674-4527/12/7/002}

\bibitem[{{Zwitter} {et~al.}(2018){Zwitter}, {Kos}, {Chiavassa}, {Buder},
  {Traven}, {{\v{C}}otar}, {Lin}, {Asplund}, {Bland-Hawthorn}, {Casey}, {De
  Silva}, {Duong}, {Freeman}, {Lind}, {Martell}, {D'Orazi}, {Schlesinger},
  {Simpson}, {Sharma}, {Zucker}, {Anguiano}, {Casagrande}, {Collet}, {Horner},
  {Ireland}, {Kafle}, {Lewis}, {Munari}, {Nataf}, {Ness}, {Nordlander},
  {Stello}, {Ting}, {Tinney}, {Watson}, {Wittenmyer}, \&
  {{\v{Z}}erjal}}]{2018MNRAS.481..645Z}
{Zwitter}, T., {Kos}, J., {Chiavassa}, A., {et~al.} 2018, \mnras, 481, 645,
  \dodoi{10.1093/mnras/sty2293}

\end{thebibliography}

\end{document}